\documentclass[useAMS,usenatbib]{mn2e}
\usepackage{graphicx}
\usepackage{rotating,times,pictex,graphicx,latexsym}
\usepackage{color}
\usepackage{longtable,amsmath}
\usepackage{lscape}
\title{Extinction techniques and impact on dust property determination}

\author[D.~Froebrich \& C.~del Burgo]{D.~Froebrich$^1$\thanks{E-mail:
df@cp.dias.ie} and C.~del Burgo$^{1,2}$ \\ $^1$ Dublin Institute for Advanced
Studies, 5 Merrion Square, Dublin 2, Ireland \\ $^2$ Dunsink Observatory,
Castleknock, Dublin 15, Ireland }

\begin{document}

\date{Received sooner; accepted later}
\pagerange{\pageref{firstpage}--\pageref{lastpage}} \pubyear{2005}
\maketitle

\label{firstpage}

\begin{abstract} 

The near infrared extinction powerlaw index ($\beta$) and its uncertainty is
derived from three different techniques based on star counts, colour excess and
a combination of them. We have applied these methods to 2MASS data to determine
maps of $\beta$ and near infrared extinction of the small cloud IC\,1396\,W. The
combination of star counts and colour excess results in the most reliable method
to determine $\beta$. It is found that the use of the correct $\beta$-map to
transform colour excess values into extinction is fundamental for column density
profile analysis of clouds. We describe how artificial photometric data, based
on the model of stellar population synthesis of the Galaxy (Robin et al.
\cite{2003A&A...409..523R}), can be used to estimate uncertainties and derive
systematic effects of the extinction methods presented here. We find that all
colour excess based extinction determination methods are subject to small but
systematic offsets, which do not affect the star counting technique. These
offsets occur since stars seen through a cloud do not represent the same
population as stars in an extinction free control field.

\end{abstract}


\section{Introduction}

Dust plays a major role in the dynamics of star formation. The determination of
its properties and distribution is hence a major challenge towards a better
understanding. There are two widely used methods (star counts and colour excess)
to determine the distribution and column density of dust in the interstellar
medium.

The original idea of using star counts to measure dust extinction dates back to
Wolf \cite{1923AN....219..109W}. Later Bok \cite{1956AJ.....61..309B} improved
the accuracy by implementing star counts at the completeness limit of the
images. The difference between the measured and intrinsic colour of a star
(colour excess) was used by Lada et al. \cite{1994ApJ...429..694L} to determine
the extinction. This approach was optimised by Lombardi \& Alves
\cite{2001A&A...377.1023L} from multi-band photometry. Recently Lombardi
\cite{2005A&A...438..169L} investigated how the two methods can be optimally
combined.

Star counts gauge extinction at a given wavelength. The colour excess requires
the knowledge of the reddening law to obtain an extinction measurement. Further
the reddening law is required to transform these values into the widely used
optical extinction ($A_{\rm V}$). This transformation law is not known {\it a
priori}. There is reasoning that a universal extinction law is valid for the
diffuse interstellar medium (Draine \cite{2003ARA&A..41..241D}). In translucent
and dense regions, however, processes like coagulation of dust grains and mantle
growth can lead to a change in the dust properties (Whittet et al.
\cite{2001ApJ...547..872W}; Cambr\'esy et al. \cite{2001A&A...375..999C}; del
Burgo et al. \cite{2003MNRAS.346..403D}; del Burgo \& Laureijs
\cite{2005MNRAS.360..901D}), which could result in a different extinction law
(Larson \& Whittet \cite{2005ApJ...623..897L}). Ultraviolet and cosmic ray
radiation has also some influence in the density and properties of dust grains
(e.g. del Burgo \& Cambr\'esy \cite{2006MNRAS.inprep.D}). Froebrich et al.
\cite{2005A&A...432L..67F} showed that the near infrared extinction powerlaw
index ($\beta$) within dark clouds can vary significantly from the value
obtained for the diffuse interstellar medium. 

The purpose of the present paper is the following. We describe an easy-to-use
and fast method to determine $\beta$ from multi-filter broad-band observations
(Sect.\,\ref{theory}). This method is applied in Sect.\,\ref{ic1396w} to an
example cloud (IC\,1396\,W). We discuss the implications of using the determined
$\beta$ instead of the standard value of 1.85 (Draine
\cite{2003ARA&A..41..241D}), for density profile analyses of dark clouds. In
Sect.\,\ref{exttest} we present a technique to test the accuracy of extinction
determination methods by artificial photometric catalogues based on the model of
stellar population synthesis of the Galaxy by Robin et al.
\cite{2003A&A...409..523R}. We finally discuss systematic uncertainties
introduced into extinction maps when using colour excess methods
(Sect.\,\ref{distcorr}). Our conclusions are then put forward in
Sect.\,\ref{conclusions}. 

\section{Determination of the NIR extinction power law index}

\label{theory}

There are several possible methods to determine the near infrared extinction
power law index ($\beta$). In this section we describe techniques based on star
counts, colour excess and a combination of both. The description is done in a
general way, in order to adapt the procedure to different datasets. We discuss
in detail the uncertainties of each method.

\subsection{General notes}

Each method discussed below will determine $\beta$ as a function of two
variables $(u, v)$. Those can be extinction and/or colour excess. Hence we can
write $\beta = f(u, v)$. Each variable consists of $N$ different values (pixels
in the extinction and/or colour excess maps). Following Taylor, we describe the
difference of the individually measured $\beta$ values from the mean value as:

\begin{equation}
\beta_i - \bar{\beta} = (u_i - \bar{u}) \frac{\partial{\beta}}{\partial{u}} +
(v_i - \bar{v}) \frac{\partial{\beta}}{\partial{v}} 
\end{equation} 
A possible estimator of the variance of $\beta$, defined as
\begin{equation}
\sigma^2_\beta \equiv \frac{1}{N - 1} \sum\limits^N_{i=1} (\beta_i -
\bar{\beta})^2 
\end{equation}
can thus be expressed by the following equation:
\begin{equation}
\sigma^2_\beta = \sigma^2_u \left(\frac{\partial{\beta}}{\partial{u}}\right)^2 +
\sigma^2_v \left(\frac{\partial{\beta}}{\partial{v}}\right)^2 + 2 \sigma_{uv}
\left(\frac{\partial{\beta}}{\partial{u}}\right)
\left(\frac{\partial{\beta}}{\partial{v}}\right).
\end{equation} 
$\sigma^2_u$ and $\sigma^2_v$ refer to the estimators of the variance of
the variables $(u, v)$. They can be estimated as the square of the noise in the
input extinction and/or colour excess maps. The estimator of the covariance of
the two input variables
\begin{equation}
\sigma_{uv} \equiv \frac{1}{N - 1} \sum\limits^N_{i=1}
(u_i-\bar{u})(v_i-\bar{v}) 
\end{equation}
can be determined from the extinction and/or colour excess maps as follows:
multiply both images and  determine the mean pixel value of the product image in
an extinction free region. Note that in an extinction free field the mean values
$\bar{u}$ and $\bar{v}$ are zero. 

In the next three subsections we introduce three different methods to
determine the function $\beta = f(u, v)$ and the resulting estimator of the
variance ($\sigma^2_\beta$).

\subsection{Star Counts}

Extinction mapping using accumulative star counts allows us to determine the
extinction at wavelength $\lambda$ where the photometry is obtained (e.g. Wolf
\cite{1923AN....219..109W}; Cambr\'esy et al. \cite{2002AJ....123.2559C}; 
Froebrich et al. \cite{2005A&A...432L..67F}). It is assumed that extinction has
been measured with two different filterbands at reference wavelengths
($\lambda_1 < \lambda_2$). Also we assume the extinction has a power-law
dependence $A_\lambda \propto \lambda^{-\beta}$ in the wavelength range of
interest. Then, the unknown index $\beta$ can be determined from:

\begin{equation} 
\label{eq1}
\beta = \frac{\ln (A_{\lambda_1} / A_{\lambda_2})}{\ln (\lambda_2 /
\lambda_1)}. 
\end{equation}
 
To estimate the uncertainty of $\beta$ we introduce two parameters: 1)
$\alpha_{\rm sc}$, which corresponds to the ratio of the estimators of the
variance in the two star count maps, i.e. $\alpha_{\rm sc} \equiv
\sigma^2_{A_{\lambda_1}} / \sigma^2_{A_{\lambda_2}}$ 2) $\gamma_{\rm sc}$, which
corresponds to the ratio of the estimator of the variance in the star count map
at $\lambda_1$ and the estimator of the covariance of the two star count maps,
i.e. $\gamma_{\rm sc} \equiv \sigma^2_{A_{\lambda_1}} / \sigma_{A_{\lambda_1}
A_{\lambda_2}}$. The estimator of the variance of $\beta$ can then be expressed
as:

\begin{equation} 
\sigma^2_\beta = \frac{1}{\ln^2\left( \lambda_2 / \lambda_1 \right)}
\frac{\sigma^2_{A_{\lambda_1}}}{A^2_{\lambda_1}} \left[ 1 + \frac{1}{\alpha_{\rm
sc}} \left( \frac{\lambda_2}{\lambda_1} \right)^{2 \beta} - \frac{2}{\gamma_{\rm
sc}} \left( \frac{\lambda_2}{\lambda_1} \right)^\beta \right]
\label{deltabeta1}
\end{equation}

 %
 %

\begin{figure*}
\includegraphics[height=8.5cm,angle=-90,bb=20 30 550 770]{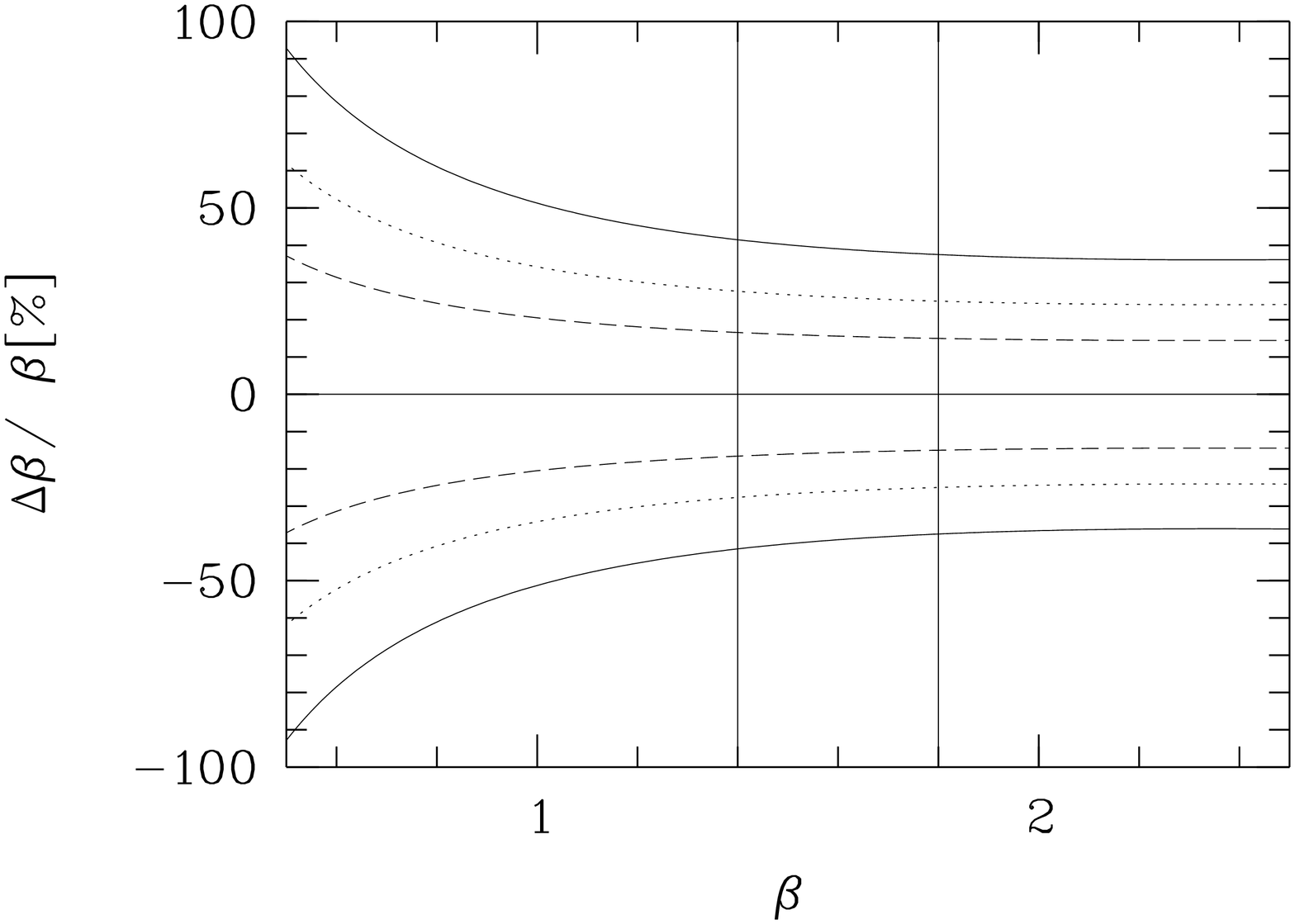}
\hfill
\includegraphics[height=8.5cm,angle=-90,bb=20 30 550 770]{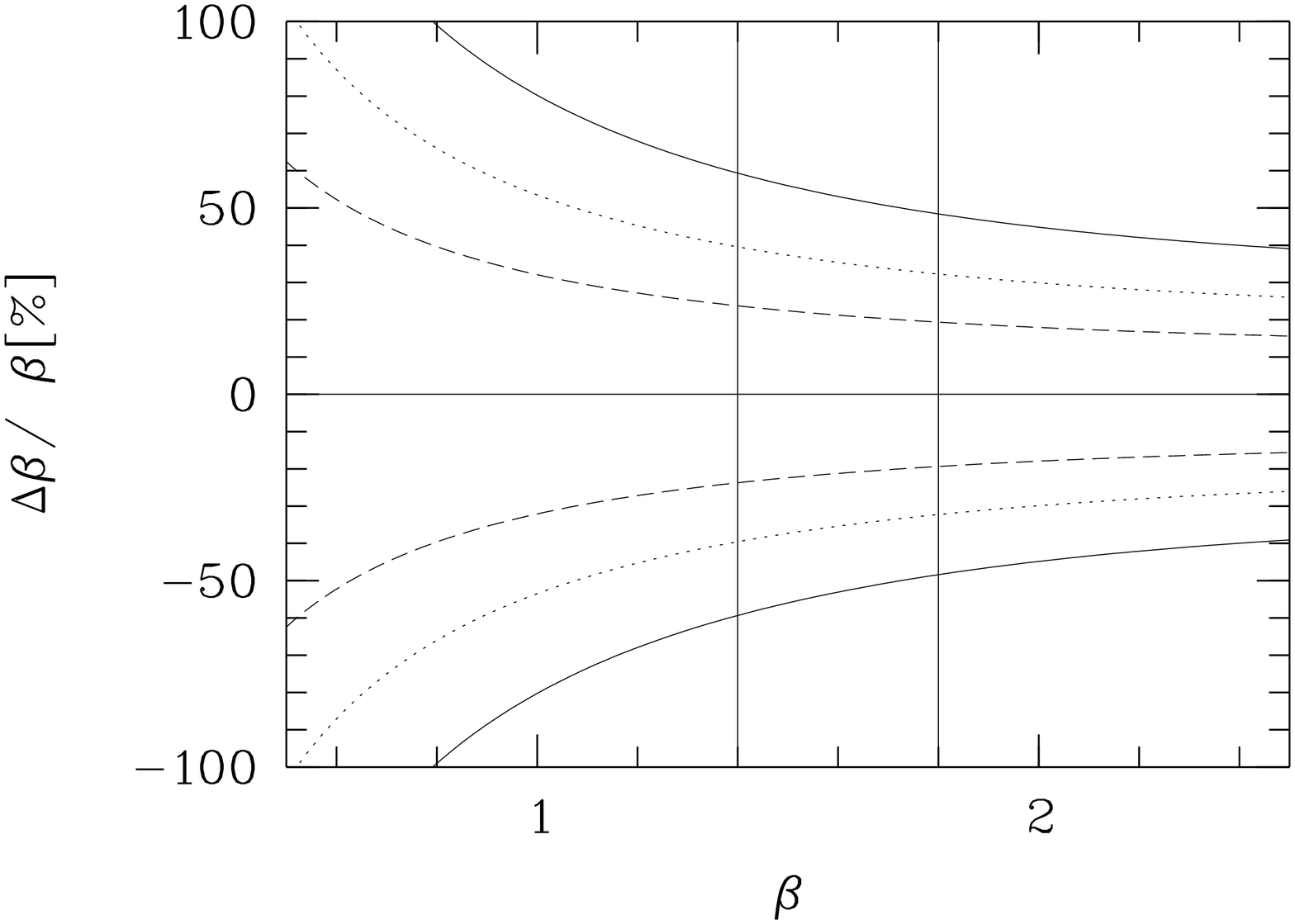}
\\
\includegraphics[height=8.5cm,angle=-90,bb=20 30 550 770]{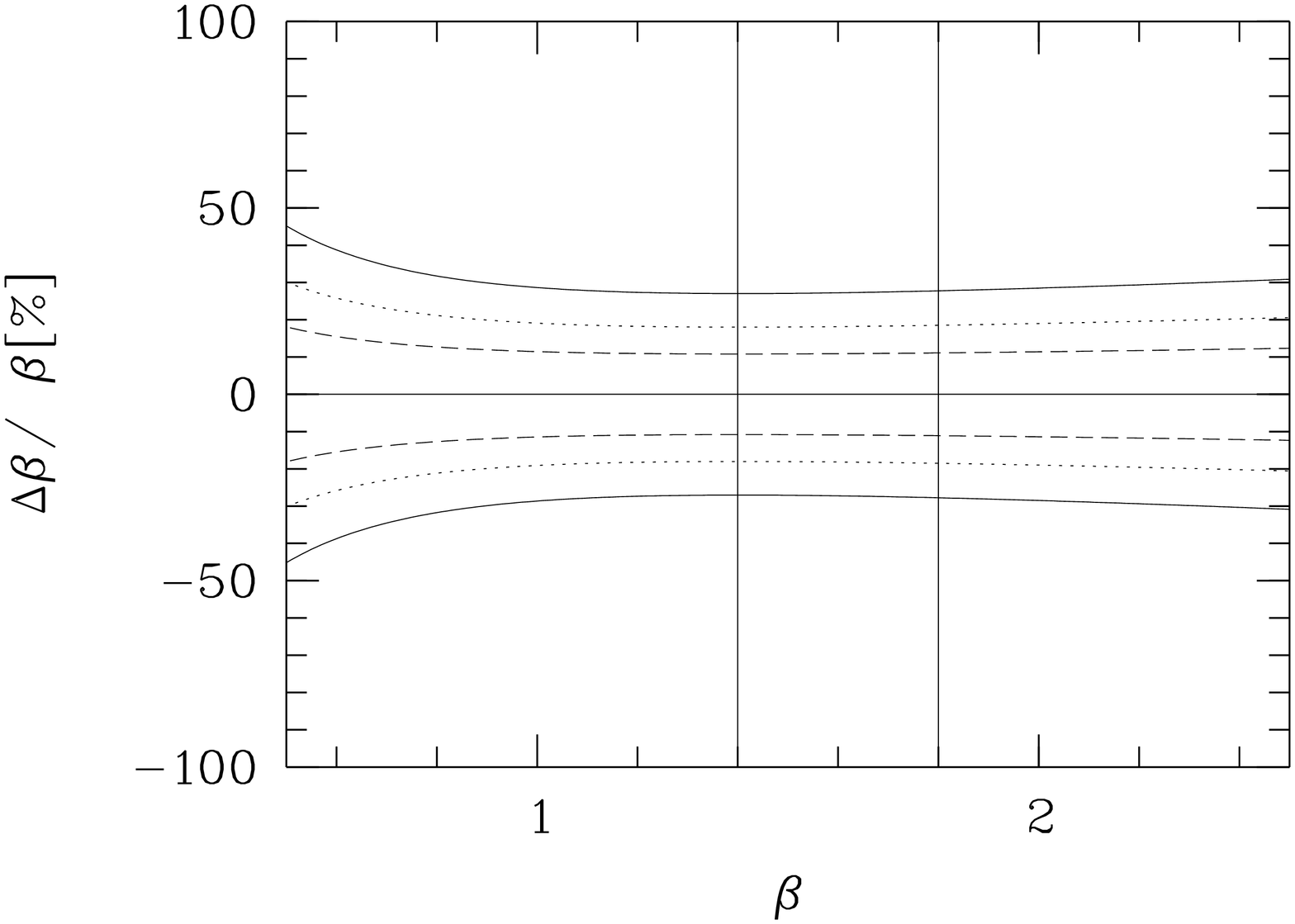}
\hfill
\includegraphics[height=8.5cm,angle=-90,bb=20 30 550 770]{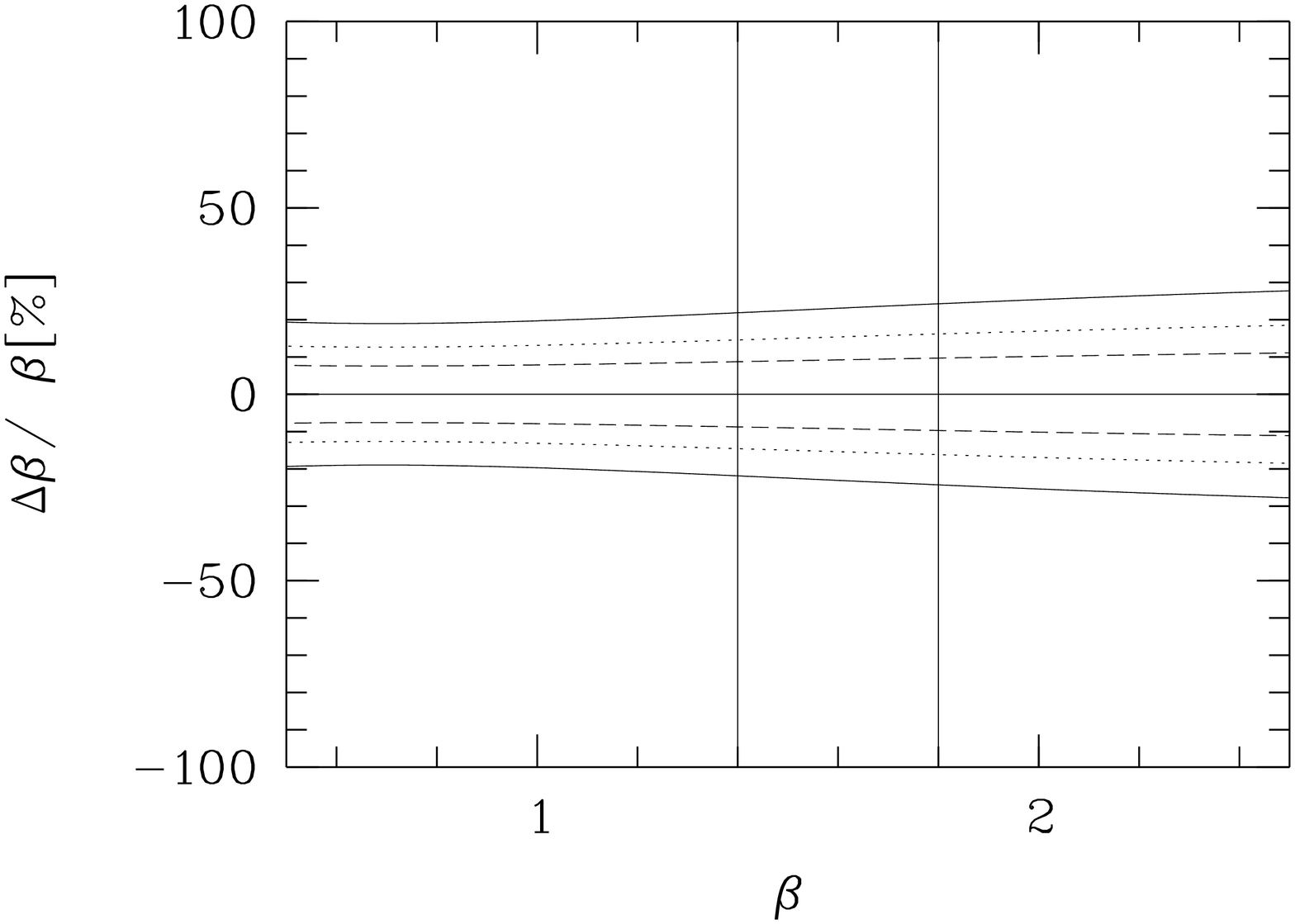}
\caption{\label{err_bet} {\bf Top-left:} Relative uncertainty of $\beta$
obtained from J- and H-band star counts. {\bf Top-right:} Relative uncertainty
of $\beta$ obtained from $\left< J-H \right>$ and $\left< H-K \right>$ colour
excess ratio. {\bf Bottom:} Relative uncertainty of $\beta$ obtained from the
combination of J-band star counts and $\left< J-H \right>$ colour excess. The
plots show $\Delta\beta / \beta$ referring to measurements of $A_{\rm J}$ (left
panels) and $\left< J-H \right>$ (right panels) with a S/N of 4 (solid line), 6
(dotted line), and 10 (dashed line). Values of $\alpha_{\rm sc} = 1.4$,
$\gamma_{\rm sc} = 1.4$, $\alpha_{\rm cosc} = 13$, $\gamma_{\rm cosc} = 6.5$,
$\alpha_{\rm co} = 3.6$ and $\gamma_{\rm co} = 3.7$ were used for the
determination (see text for details). The two vertical lines in the plots
indicate values of $\beta = 1.4$ and 1.8, which represent the value for dense
regions and the diffuse ISM, respectively. Note that the two bottom panels
are identical despite showing apparently different behaviour. The relative
uncertainty of $\beta$ as a function of the S/N in the $A_{\rm J}$ map
(Eq.\,\ref{eq11}) can be converted into the relative uncertainty of $\beta$ as a
function of the S/N in the $\left< J-H \right>$ map using Eq.\,\ref{eq10} and
the values of $\alpha_{\rm cosc}$ and $\gamma_{\rm cosc}$.}
\end{figure*}

Figure\,\ref{err_bet} (top-left) shows, as an example, the relative uncertainty
$\Delta\beta / \beta = \sqrt{\sigma^2_\beta} / \beta$ versus $\beta$ for typical
star counts in J- and H-band 2MASS data. The three different lines in
Fig.\,\ref{err_bet} (top-left) denote extinction measurements $A_{\rm J}$ of 4,
6 and 10$\sigma$. For 2MASS data in the area of IC\,1396\,W we have determined
$\alpha_{\rm sc} = 1.4$ and $\gamma_{\rm sc} = 1.4$ (see below). It is found
that a high accuracy for the extinction measurements is required to obtain a
good signal to noise ratio (S/N) for $\beta$; for typical values of $\beta =
1.4$ at least a $4 \sigma$ detection of the extinction in the J-band is needed
to determine $\beta$ with an accuracy of 40\,\%. Note that this is valid for a
single measurement only (1 independent pixel in the extinction map). The S/N can
be improved by averaging over several independent measurements (see below).

\subsection{Colour excess}

The most simple approach of a colour excess method is to measure the apparent
brightness of a star at two different wavelengths, $m_{\lambda_1}$ and
$m_{\lambda_2}$, and compare the colour $m_{\lambda_1}$ - $m_{\lambda_2}$ with
the theoretical value for main sequence stars or the colour of stars measured in
an extinction free control field $m^{\rm tr}_{\lambda_1}$ - $m^{\rm
tr}_{\lambda_2}$ (e.g. Lada et al. \cite{1994ApJ...429..694L}). The colour
excess $\left< \lambda_1 - \lambda_2 \right>$ is defined as: 

\begin{equation} 
\left< \lambda_1 - \lambda_2 \right> \equiv \left( m_{\lambda_1} - m_{\lambda_2}
\right) - \left( m^{\rm tr}_{\lambda_1} - m^{\rm tr}_{\lambda_2} \right) 
\end{equation}

If we assume that this colour excess is entirely due to extinction and, as for
the star count analysis, a constant $\beta$ is valid, we can determine the
extinction at $\lambda_2$ by

\begin{equation} 
A_{\lambda_2} = \frac{\left< \lambda_1 - \lambda_2 \right>}{ \left(
\frac{\lambda_2}{\lambda_1} \right)^{\beta} - 1}.
\end{equation}
 
Measuring $\left< \lambda_2 - \lambda_3 \right>$, with $\lambda_1 < \lambda_2 <
\lambda_3$, we as well can determine the extinction at $\lambda_2$:

\begin{equation} 
A_{\lambda_2} = \frac{\left< \lambda_2 - \lambda_3 \right>}{ 1 - \left(
\frac{\lambda_3}{\lambda_2} \right)^{-\beta}}
\end{equation}

The right hand side of both equations should essentially give the same value. We
can define the colour excess ratio $R \equiv \frac{\left< \lambda_1 - \lambda_2
\right>}{\left< \lambda_2 - \lambda_3 \right>}$ and obtain the following
equation for $\beta$:

\begin{equation} 
0 = \left( \frac{\lambda_2}{\lambda_1} \right)^{\beta} + R \cdot \left(
\frac{\lambda_3}{\lambda_2} \right)^{-\beta} - R - 1
\label{betacolour}
\end{equation}

\begin{figure}
\includegraphics[height=8.cm,angle=-90,bb=20 30 550 770]{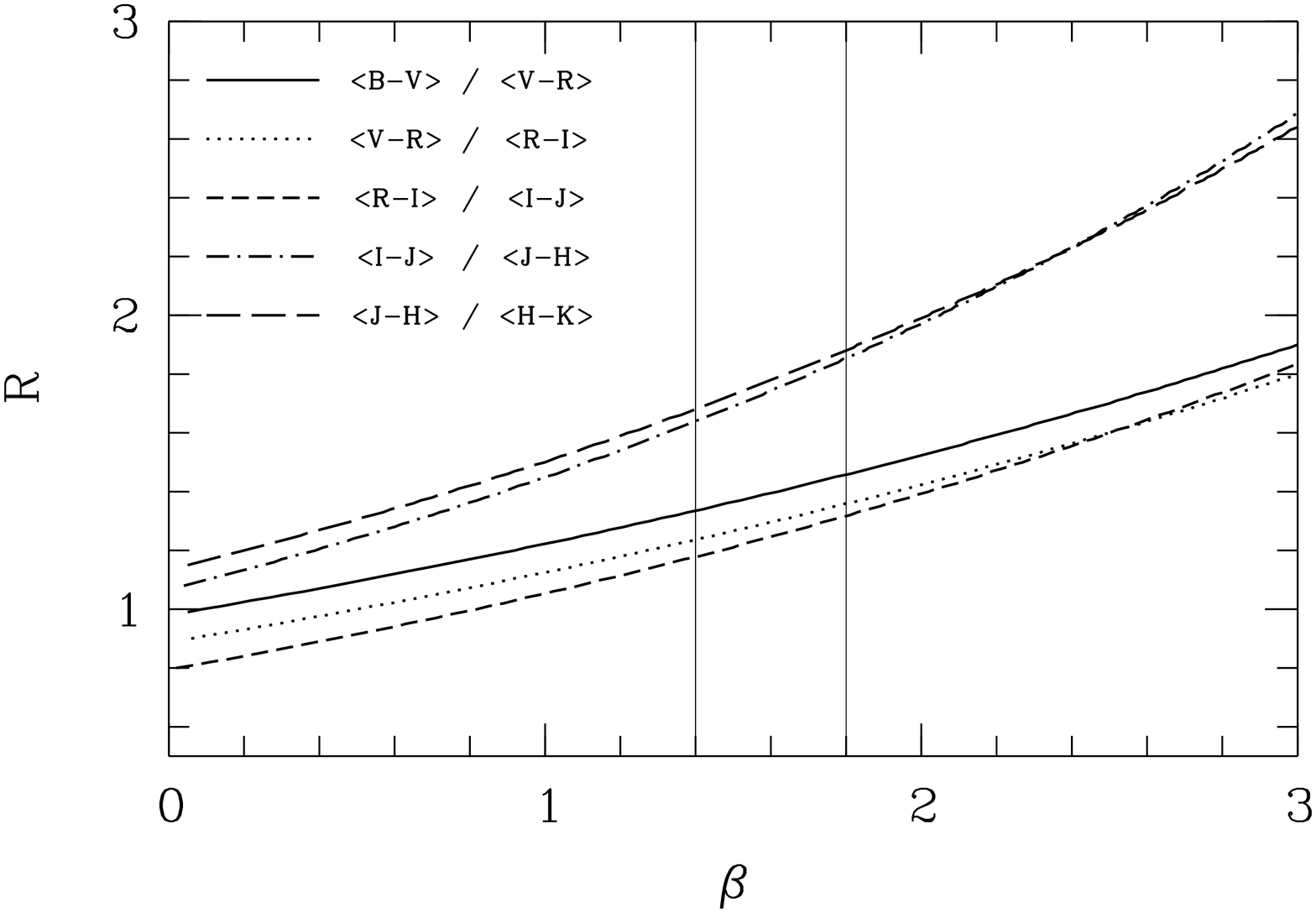}
\caption{\label{bet_r} Solution of Eq.\,\ref{betacolour} for a number of common
filter/wavelength combinations. This plot allows us to estimate $\beta$ from the
ratio of the colour excess of the two shorter and the two longer wavelength. We
plot: BVR (solid), VRI (dotted), RIJ (short-dash), IJH (dash-dot) and JHK
(long-dash).}    
\end{figure}

Fig.\,\ref{bet_r} shows the solution of Eq.\,\ref{betacolour} for some filter
combinations. These plots allow us to estimate $\beta$ for a certain filterband
combination and measured colour excess ratio. 

Since $\beta$ cannot be determined analytically from Eq.\,\ref{betacolour}, we
assume that $\beta = {\bf B}(R)$ provides a solution of Eq.\,\ref{betacolour}
with sufficient accuracy (the function ${\bf B}(R)$ could be a polynomial of
order $n$). Then we can determine the estimator of the variance of $\beta$
using:

\begin{equation} 
\sigma^2_\beta = \frac{\sigma^2_{\left< \lambda_1 - \lambda_2 \right>}}{\left<
\lambda_1 - \lambda_2 \right>^2} R^2 \left[ 1 + \left( \frac{R}{\alpha_{\rm co}}
\right)^2 - 2 \frac{R}{\gamma_{\rm co}} \right] \left( \frac{\partial {\bf
B}(R)}{\partial R} \right)^2 
\label{deltabetacol1}
\end{equation}

 %
 %

Note that the partial derivative $\partial {\bf B}(R) / \partial R$ can
directly be obtained from Eq.\,\ref{betacolour} by using the inverse function
theorem. We define the parameters $\alpha_{\rm co} \equiv \sigma^2_{\left<
\lambda_1 - \lambda_2 \right>} / \sigma^2_{\left< \lambda_2 - \lambda_3
\right>}$ and $\gamma_{\rm co} \equiv \sigma^2_{\left< \lambda_1 - \lambda_2
\right>} / \sigma_{\left< \lambda_1 - \lambda_2 \right>\left< \lambda_2 -
\lambda_3 \right>}$. Average values based on 2MASS data for IC\,1396\,W are
$\alpha_{\rm co} = 3.6$ and $\gamma_{\rm co} = 3.7$. Note that these values
depend on the intrinsic scatter of the stellar colours, the spacial distribution
of stars and the photometric uncertainties in the data, and are hence subject to
change slightly with position. In Fig.\,\ref{err_bet} (top-right) the resulting
relative uncertainty of $\beta$ for this example is plotted. Similar to the star
counting technique, high S/N colour excess maps are required to reliably
estimate $\beta$. We note, however, that for the same dataset usually a higher
S/N is achieved for the colour excess method than for the star counting (see
e.g. Cambr\'esy et al \cite{2002AJ....123.2559C}). 

\subsection{Combining star counts and colour excess}

If observations at only two wavelengths $\lambda_1 < \lambda_2$ are available,
we do not necessarily need to determine $\beta$ solely from star counts. We can
combine star counts and colour excess. If we measure the extinction
$A_{\lambda_1}$ by star counts, the colour excess $\left< \lambda_1 - \lambda_2
\right>$, and assume that $\beta$ is constant in the wavelength range of
interest, then $\beta$ can be determined as:

\begin{equation}
\beta = - \ln \left( 1 - \frac{\left< \lambda_1 - \lambda_2
\right>}{A_{\lambda_1}} \right) / \ln \left( \lambda_2 / \lambda_1 \right)
\label{eq10}
\end{equation}

 %
 %

As for the other two methods, we can estimate the uncertainty associated to the
procedure. Similar to those we define the parameters $\alpha_{\rm cosc} \equiv
\sigma^2_{A_{\lambda_1}} / \sigma^2_{\left< \lambda_1 - \lambda_2 \right>}$ and
$\gamma_{\rm cosc} \equiv \sigma^2_{A_{\lambda_1}} / \sigma_{A_{\lambda_1}
\left< \lambda_1 - \lambda_2 \right>}$, and obtain the estimator of the
variance of $\beta$:

\begin{equation}
\begin{gathered}
\sigma^2_\beta = \frac{1}{\ln^2 \left( \lambda_2 / \lambda_1 \right)}
\frac{\sigma^2_{A_{\lambda_1}}}{A^2_{\lambda_1}} \left(
\frac{\lambda_2}{\lambda_1} \right)^{2 \beta} \hspace{3.7cm} \\ \hspace{0.25cm}
\cdot \left\{ \left[ 1 - \left( \frac{\lambda_2}{\lambda_1} \right)^{-\beta}
\right]^2 + \frac{1}{\alpha_{\rm cosc}} - \frac{2}{\gamma_{\rm cosc}} \left[ 1
- \left( \frac{\lambda_2}{\lambda_1} \right)^{-\beta} \right] \right\}
\label{eq11}
\end{gathered}
\end{equation}

 %

Figure\,\ref{err_bet} (bottom) shows two examples of the relative uncertainty of
$\beta$ when determined from the combination of star counts and colour excess.
Both cases combine J-band star counts and $\left< J-H \right>$ colour excess. In
Fig.\,\ref{err_bet} (bottom left) J-band star counts are used as a measure of
the S/N. In Fig.\,\ref{err_bet} (bottom right) the $\left< J-H \right>$ colour
excess is used. Values of $\alpha_{\rm cosc} = 13$ and $\gamma_{\rm cosc} = 6.5$
are obtained for the example (IC\,1396\,W).

\subsection{Other combinations}

If observations at three wavelengths are available and we assume that there is a
constant $\beta$, in principle a variety of other possible combinations of
colour excess and star counts emerges. One can combine colour excess between two
of the wavelengths with star counts at a third wavelength. Also two other
combinations using purely colour excess are possible. Since the procedure for
the determination of $\beta$ and its uncertainty, as well as the results are
very similar to the techniques described in the above paragraphs, we refrain
from presenting them here.

\subsection{Usage of broad-band filters}

The analysis so far is implicitly carried out using the simplifying assumption
that extinction and colour excess measurements are carried out at monochromatic
wavelengths. Usually the observations are, however, obtained using broadband
filters. Does this influence our results? 

Firstly we note that extinction and colour excess measurements are obtained by
averaging data over a number stars in a certain area. For example the colour
excess maps of IC\,1396\,W (presented below) are determined using on average 25
stars at each position. The average spectral energy distribution of these stars
indeed will change the reference wavelength of the used filter. For NIR
observations, as used here, the reference wavelength will be shorter as the
reference wavelength of the used filter. Second we note that all determination
procedures for $\beta$ and its uncertainty depend on the ratio of two
wavelength. A change of this ratio would imply systematic off-sets in our
determined $\beta$-value. Since for all NIR filters the shift in reference
wavelength is toward shorter wavelength, the resulting change in the wavelength
ratio will be small. 

To estimate this we have convolved black body curves with temperatures between
2000\,K and 12000\,K with 2MASS filter profiles. It is found that the resulting
shift in the reference wavelength ratio is at most 0.5\,\%. This generally
results in about 3\,\% systematic off-sets for the value of $\beta$. It is hence
much smaller than the determined statistical noise, which is in the order of
30\,\% or higher. Thus, the use of broad band filters does not influence our
results.

\subsection{Discussion}

In general all methods described here have similar uncertainties when applied to
the same wavelength range. But it is possible to determine the method
leading to the highest S/N in $\beta$ for any particular dataset. However, there
might be additional systematic uncertainties (see Sect.\,\ref{distcorr} for a
discussion). Figure\,\ref{err_bet} allows us to compare the different methods.
In the left panels the relative uncertainties of $\beta$ are plotted as a
function of the S/N in the J-band star count map. Note that for purely star
counting the uncertainties are larger as for the combination of star counts and
colour excess. In the right panels of Fig.\,\ref{err_bet} the relative
uncertainties of $\beta$ as a function of the S/N in the $\left< J-H \right>$
colour excess map are plotted. We find that the scatter is smaller when
combining colour excess and star counts instead of using the colour excess
ratio. This is especially the case towards smaller values of $\beta$. We
conclude that for our example (IC\,1396\,W) the combination of J-band star
counts and $\left< J-H \right>$ colour excess provides the best $\beta$-map.
Note that this might change for other datasets, since the estimator of the
variance of $\beta$ depends on the values of $\alpha_{\rm i}$ and $\gamma_{\rm
i}$.

Even when using the best method to determine $\beta$, the uncertainties based on
individual measurements (one independent pixel in the extinction/colour excess
maps) are rather large even for high S/N extinction data. However, several
pixels may be averaged to decrease the uncertainties. Although a simple
averaging over adjoining pixels would decrease the spatial resolution it is in
general a good approach in regions were no apparent changes of $\beta$ are seen
in the data (i.e. the scatter of $\beta$-values is consistent with the
statistical expectations; 68\,\% of the data points are within the 1$\sigma$
error bars, etc.). If a dependency of $\beta$ on the column density is found (as
for IC\,1396\,W, see below), we suggest to apply the following method to
determine how $\beta$ depends on the column density within individual clouds. 

\begin{figure*}
\includegraphics[height=5.8cm,angle=-90,bb=20 30 550 770]{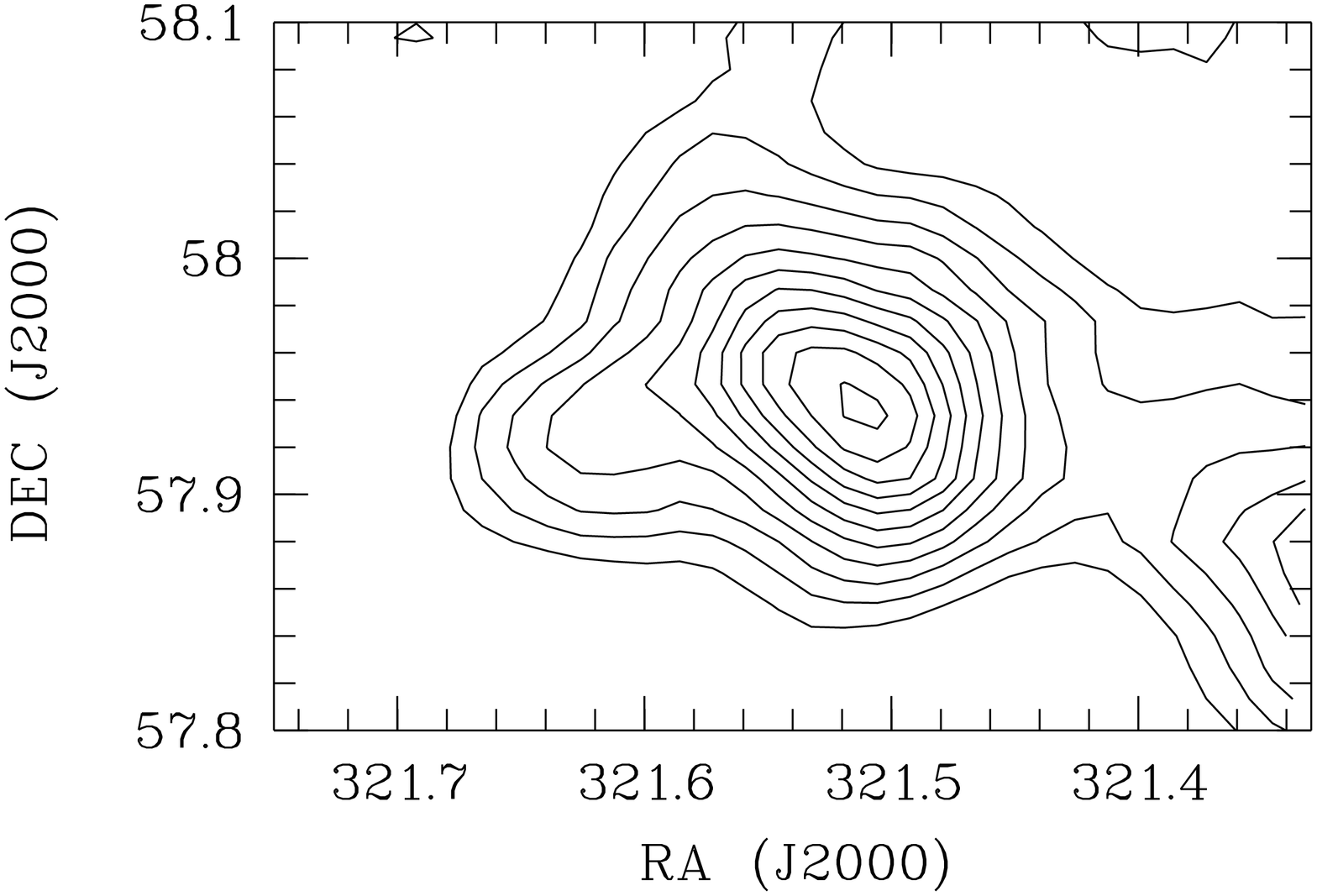}
\hfill
\includegraphics[height=5.8cm,angle=-90,bb=20 30 550 770]{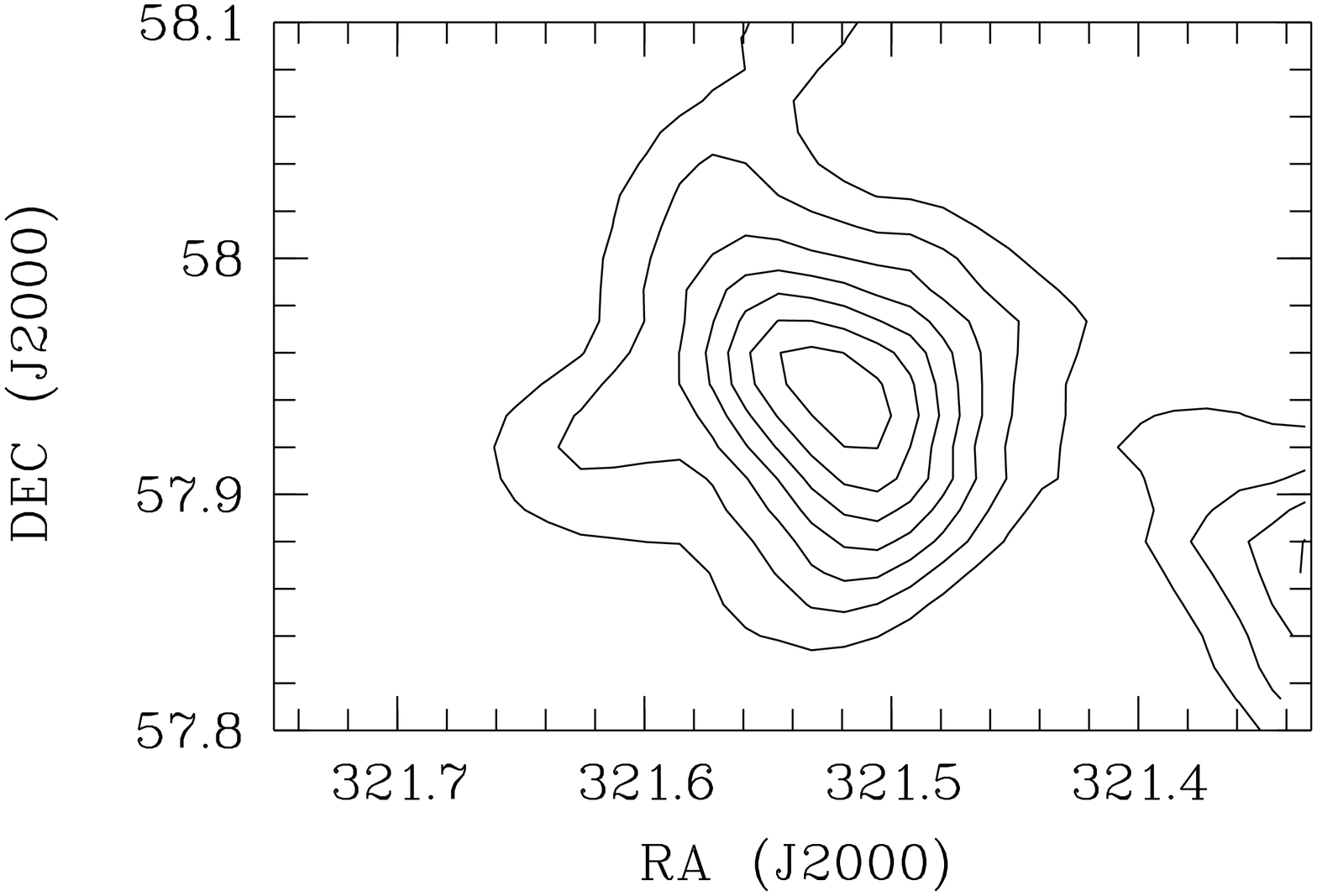}
\hfill
\includegraphics[height=5.8cm,angle=-90,bb=20 30 550 770]{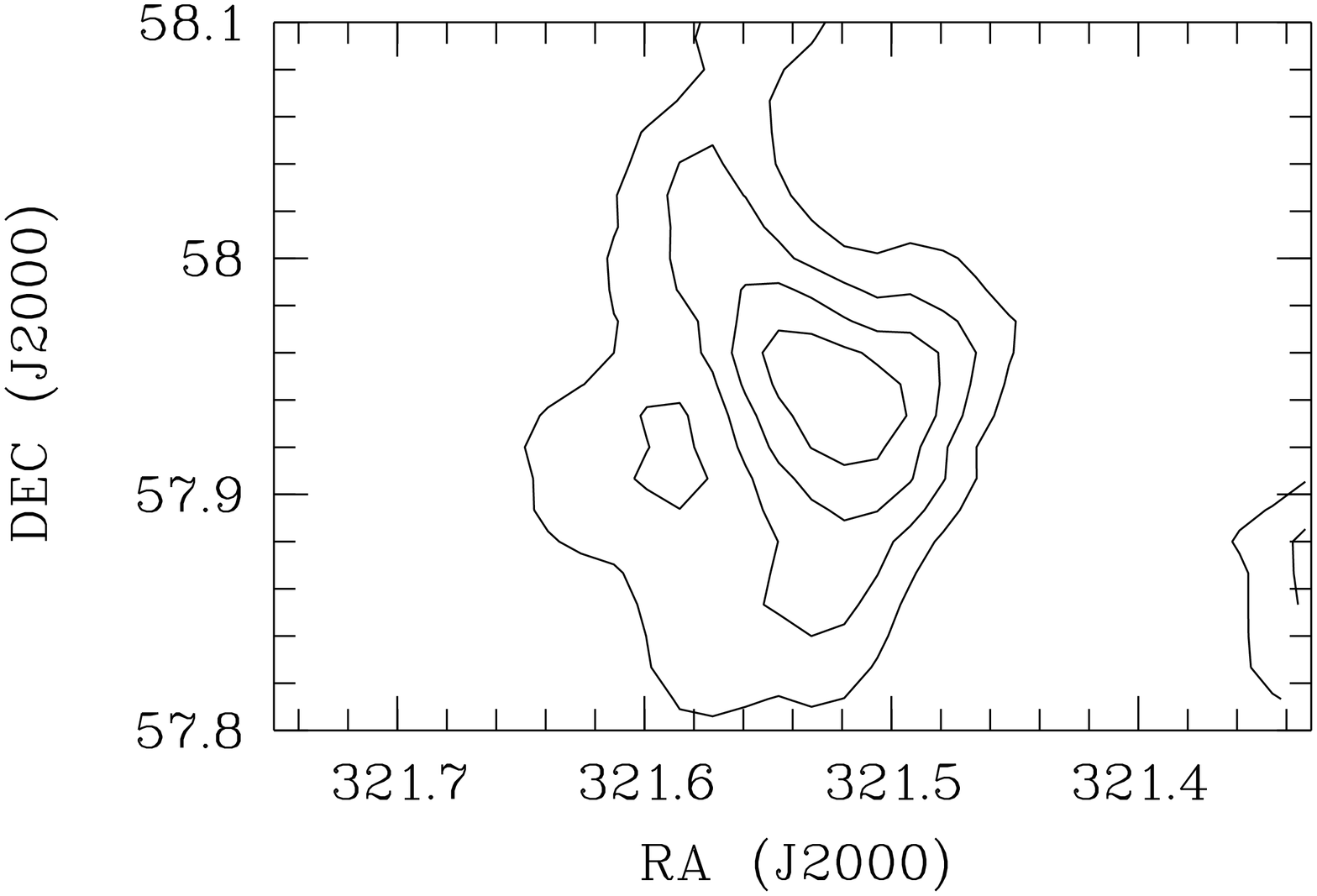}
\\
\includegraphics[height=5.8cm,angle=-90,bb=20 30 550 770]{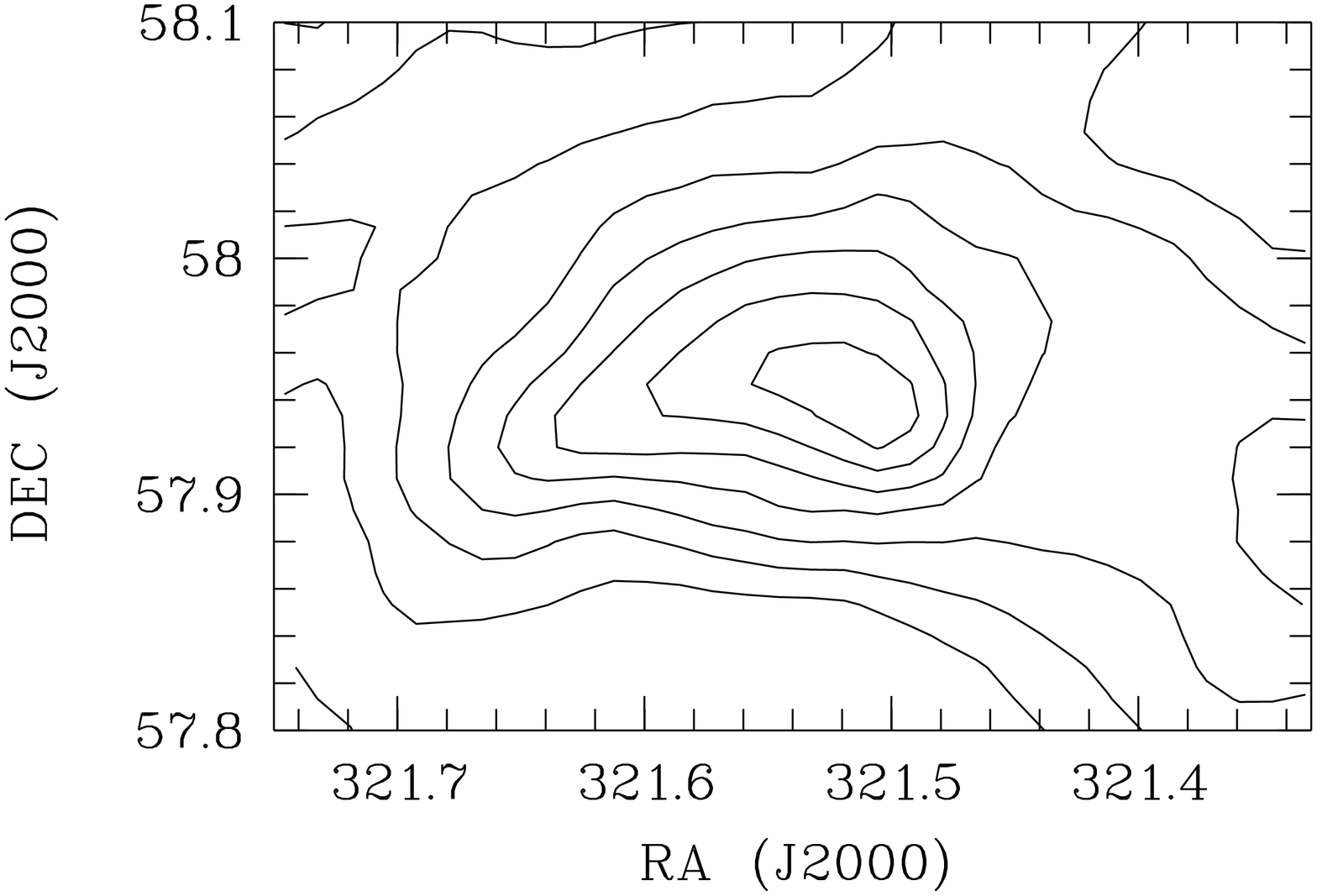}
\hfill
\includegraphics[height=5.8cm,angle=-90,bb=20 30 550 770]{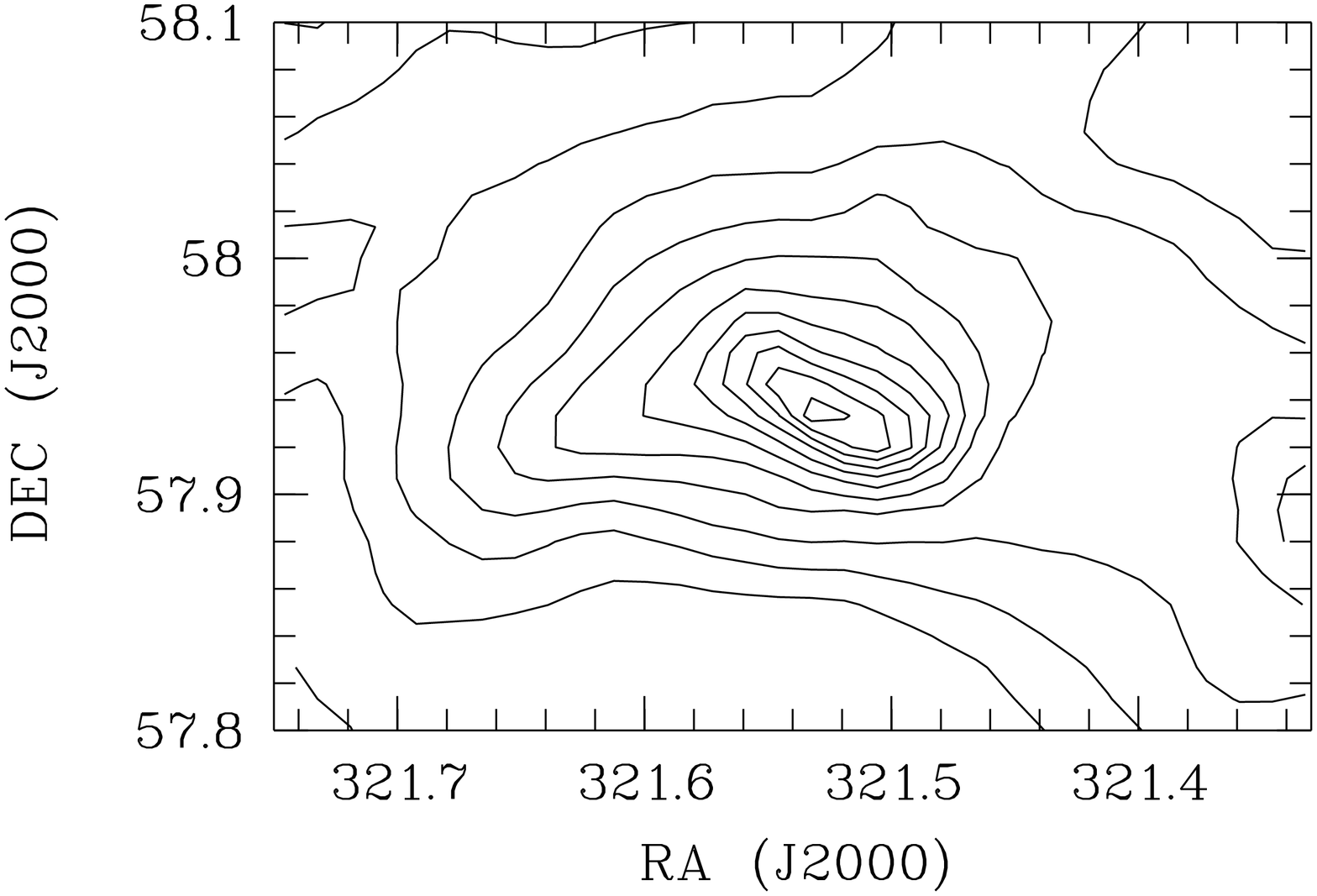}
\hfill
\includegraphics[height=5.8cm,angle=-90,bb=20 30 550 770]{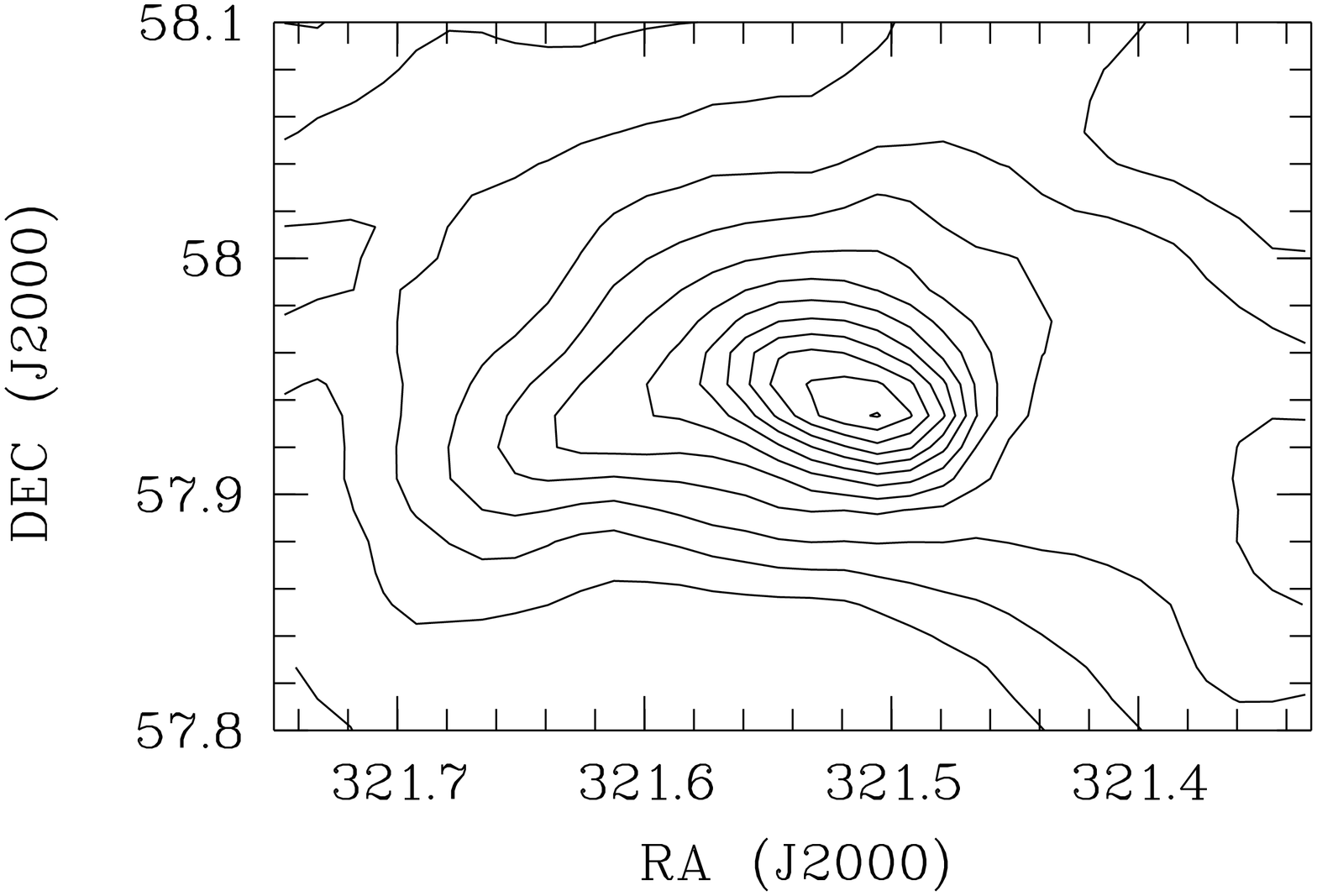}
\\
\includegraphics[height=5.8cm,angle=-90,bb=20 30 550 770]{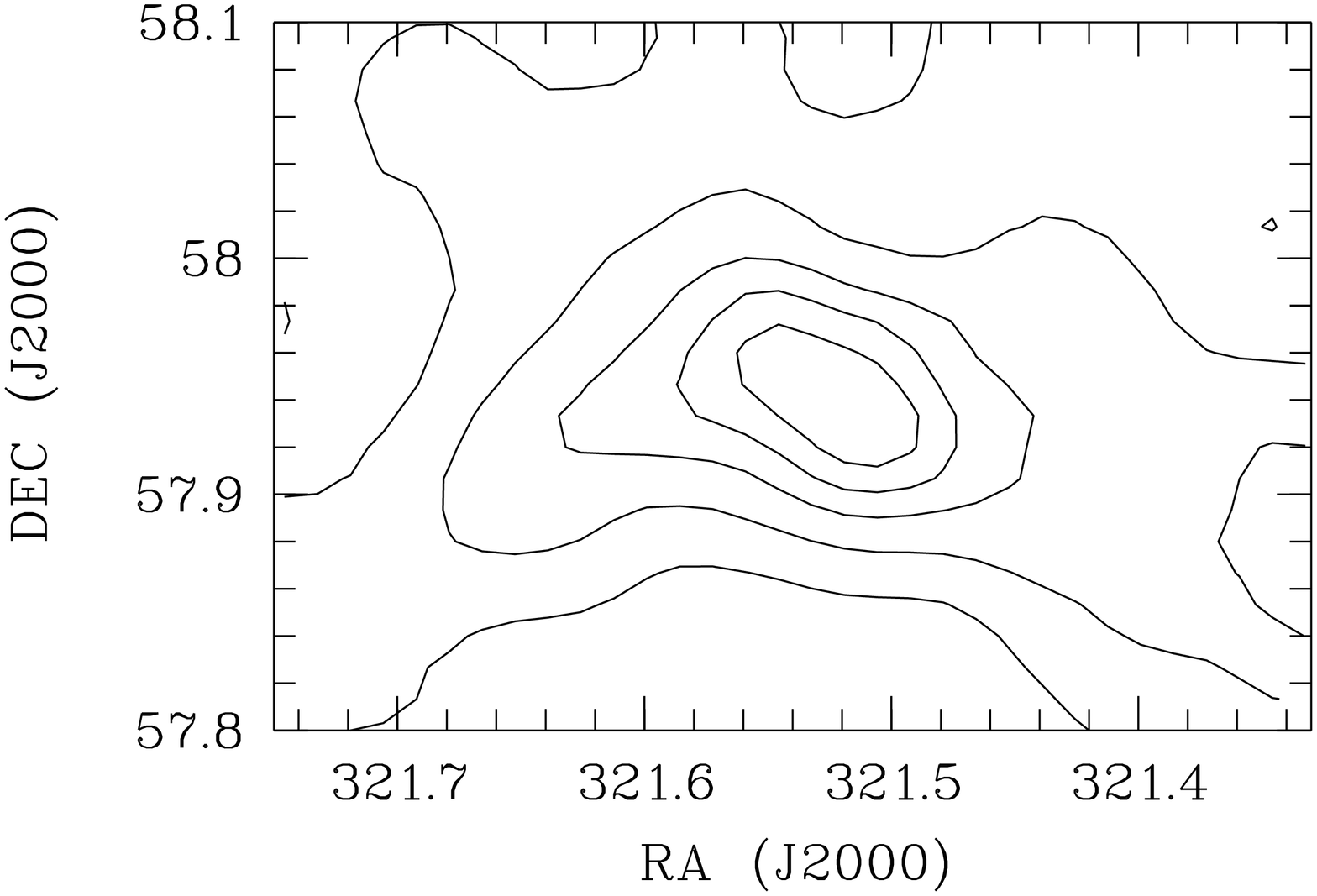}
\hfill
\includegraphics[height=5.8cm,angle=-90,bb=20 30 550 770]{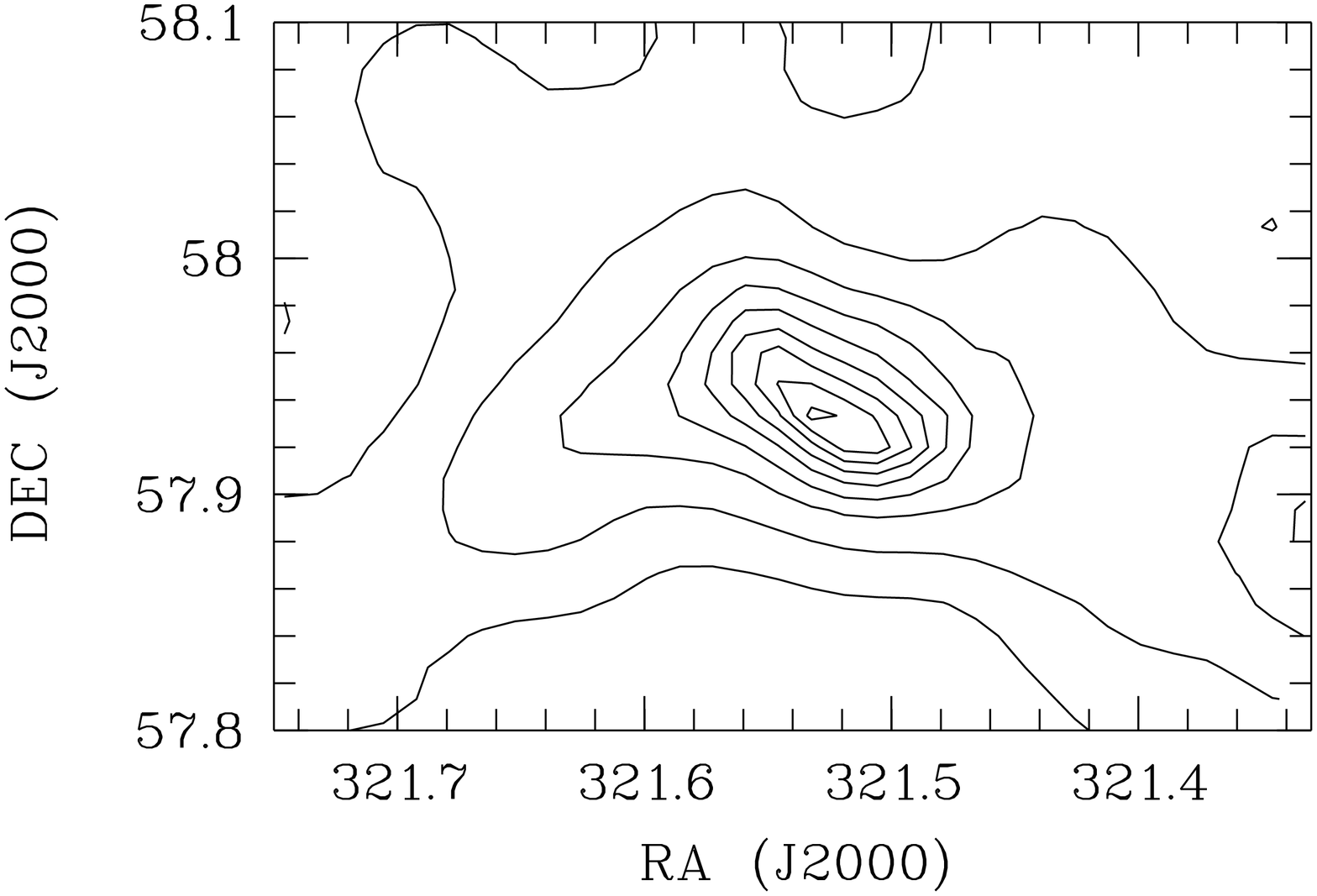}
\hfill
\includegraphics[height=5.8cm,angle=-90,bb=20 30 550 770]{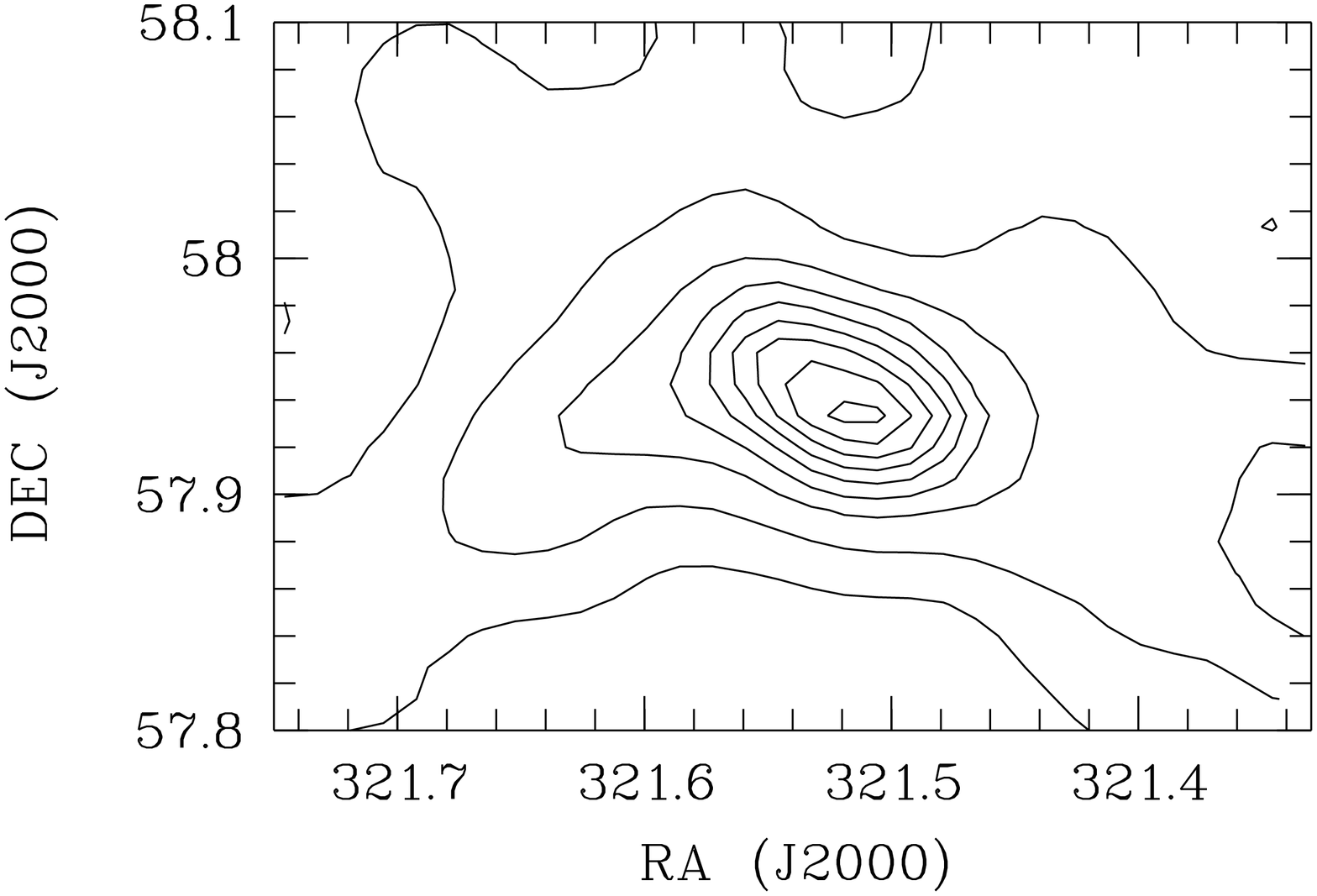}
\caption{\label{ic1396w_extmaps} {\bf Top:} Extinction maps of IC\,1396\,W
obtained from accumulative star counts in 2MASS data. Contours in each panel
start at 0.15\,mag of extinction in the J-band (left), H-band (middle) and
K-band (right) and increase by 0.15\,mag with each step. {\bf Middle:}
Extinction maps of IC\,1396\,W obtained from $\left< J-H \right>$ colour excess
in 2MASS data. Contours start at 0.15\,mag of extinction in the J-band and
increase by 0.15\,mag with each step. The conversion of colour excess into
extinction uses $\beta = 1.8$ (left), the $\beta$-map obtained from the $\left<
J-H \right>$ and $\left< H-K \right>$ colour excess ratio (middle) and the
best $\beta$-map determined from $\left< J-H \right>$ colour excess and
J-band star counts. {\bf Bottom:} The same as the middle row but for $\left< H-K
\right>$ colour excess. Contours are in H-band extinction.}
\end{figure*}

\begin{itemize}

\item {\bf Accumulated star counts or its combination with colour excess:} For
each independent pixel $\beta$ is plotted against extinction (e.g.
$A_{\lambda_1}$). A running mean in this diagram can then be used as
representation of the relation $\beta$-$A_{\lambda_1}$. Alternatively mean
values of $\beta$ in certain $A_{\lambda_1}$-bins can be fit by an appropriate
function to establish the relation $\beta$-$A_{\lambda_1}$.

\item {\bf Colour excess:} The determination of the extinction from colour
excess requires the knowledge of $\beta$. Thus, as a first step for each
independent pixel the value of $\beta$ is plotted against the colour excess
(e.g. $\left< \lambda_1-\lambda_2 \right>$). Similar to the accumulated star
count method we establish the relation $\beta$-$\left< \lambda_1-\lambda_2
\right>$ using a running mean or by fitting an appropriate function. This
relation then can be used to determine the extinction from the colour excess and
to determine e.g. the relation $\beta$-$A_{\lambda_1}$.

\end{itemize}

\section{A case study: IC\,1396\,W}

\label{ic1396w}

In order to illustrate the $\beta$-determination procedure and its implications
on extinction and cloud column density profiles, we present in this section an
example, the small cloud IC\,1396\,W. This object is part of the Cep\,OB\,2
association, about 8' in size and at roughly 750\,pc distance (e.g. Froebrich \&
Scholz \cite{2003A&A...407..207F}). The position of the cloud in the sky and the
2MASS detection limits facilitate our analysis in the way that there is only a
negligible fraction of foreground stars to the cloud. Hence we do not need to
remove foreground stars when determining extinction maps.

\subsection{Extinction maps}

In the top panel of Fig.\,\ref{ic1396w_extmaps} we show the three extinction
maps ($A_{\rm J}$, $A_{\rm H}$, $A_{\rm K}$) for this cloud obtained from
accumulative star counts. The contours start at 0.15\,mag extinction at the
filter wavelength and increase by this amount. The basic structure (central peak
and small extension to the East) of the cloud is the same in all three filters.
The extinction maps obtained from the colour excess maps show a slightly
different structure (see bottom panel of Fig.\,\ref{ic1396w_extmaps}). In
particular we can trace much better the outer areas of the cloud. There is an
apparent small offset between the star count and colour excess extinction maps.
This can be explained by possible differences in the selection of the control
field (for the colour excess an apparently IRAS emission free field south of the
cloud was chosen, in contrast to the surrounding 1$^\circ$x1$^\circ$ field for
the star counts). 

\begin{figure*}
\includegraphics[width=5.8cm,bb=0 0 550 430]{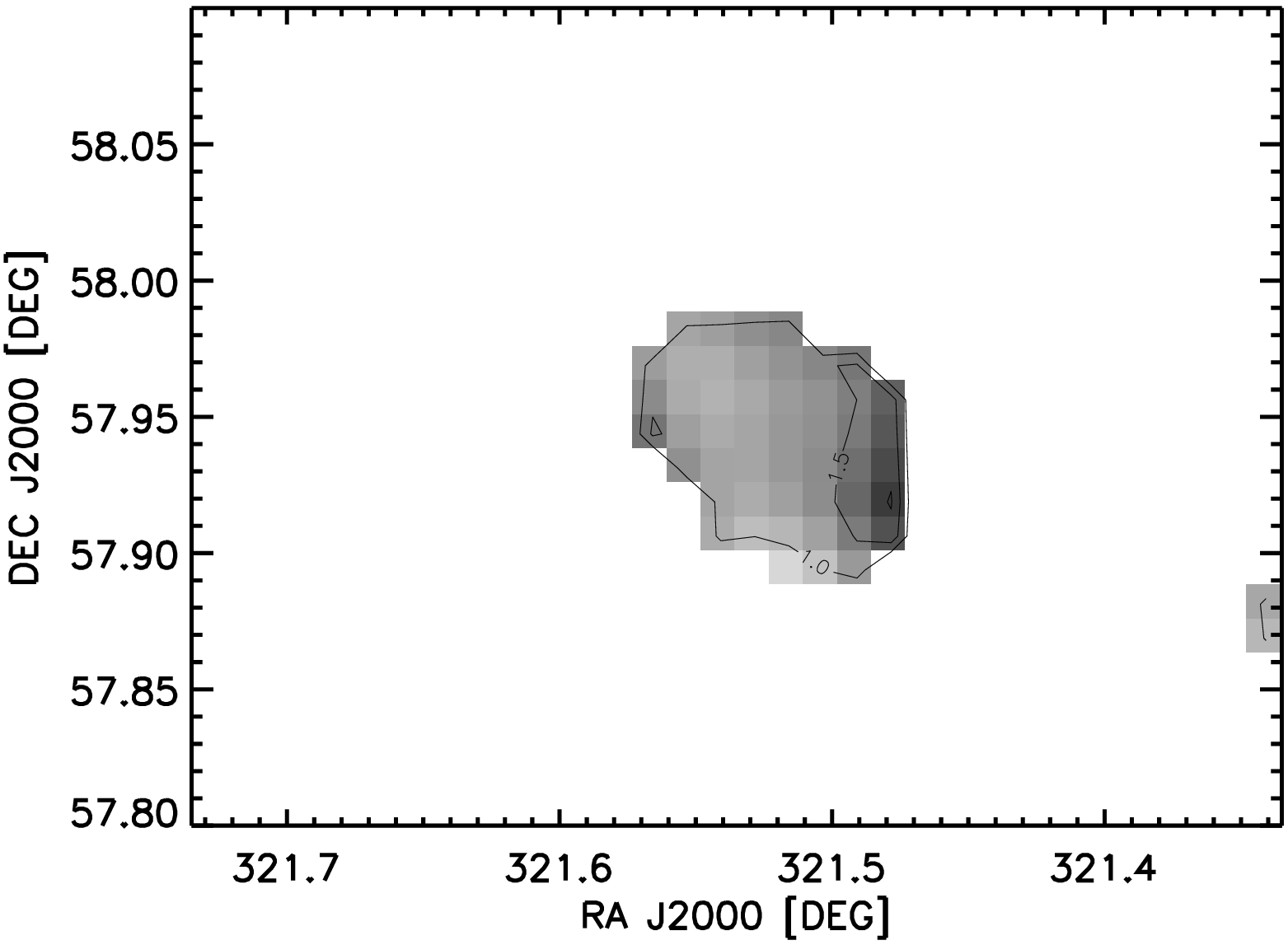}
\hfill
\includegraphics[width=5.8cm,bb=0 0 550 430]{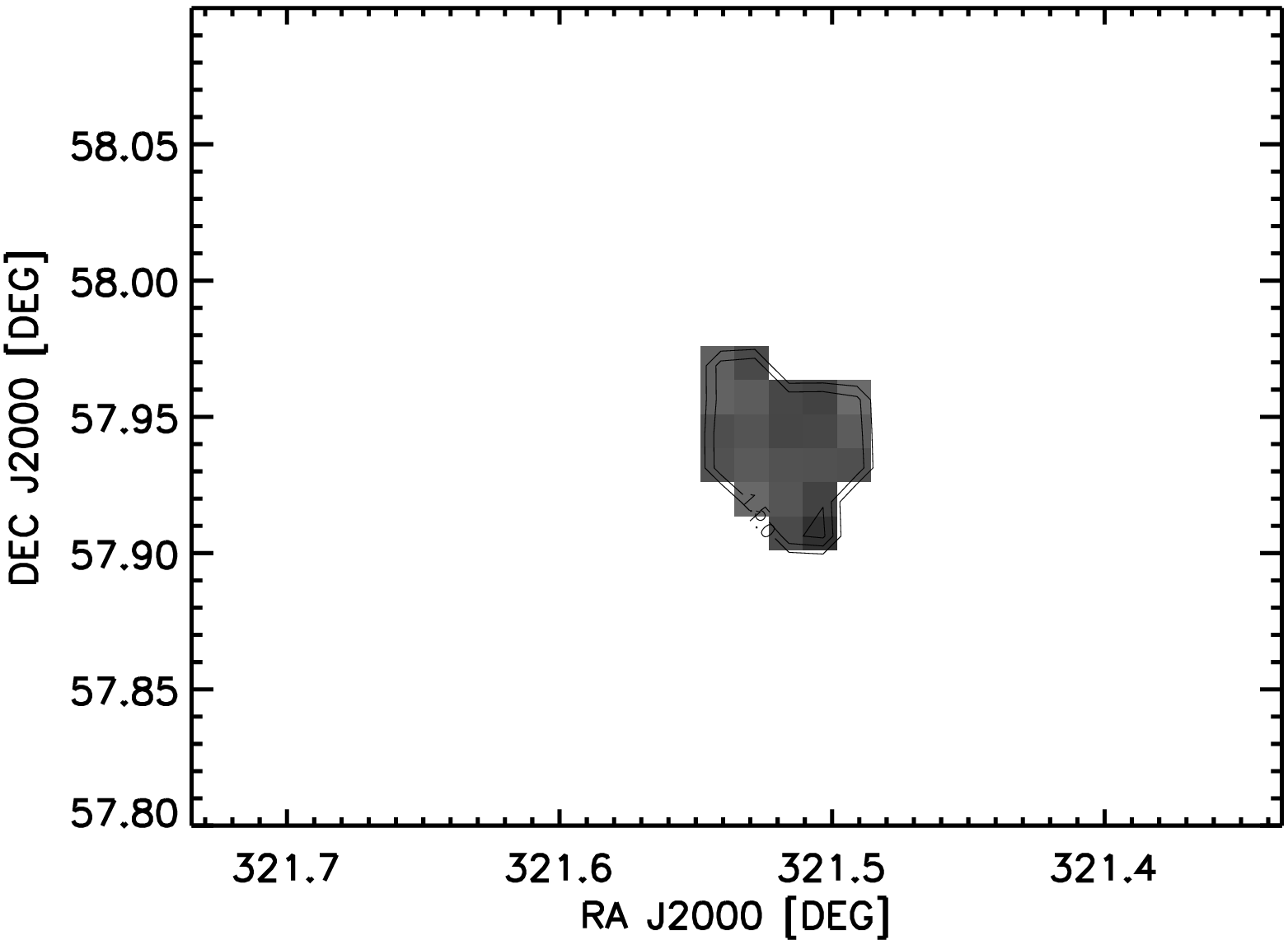}
\hfill
\includegraphics[width=5.8cm,bb=0 0 550 430]{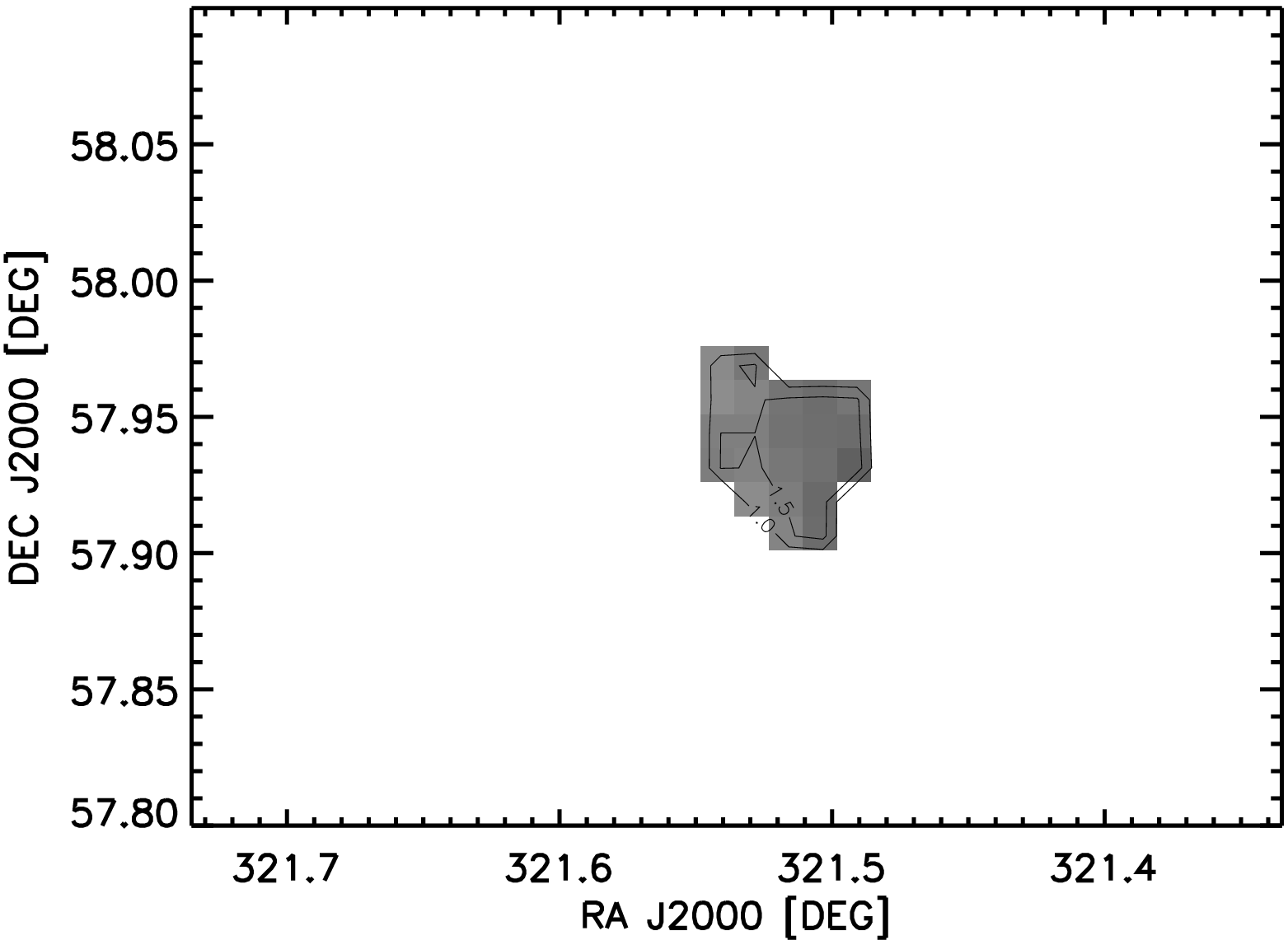}
\\
\includegraphics[width=5.8cm,bb=0 0 550 430]{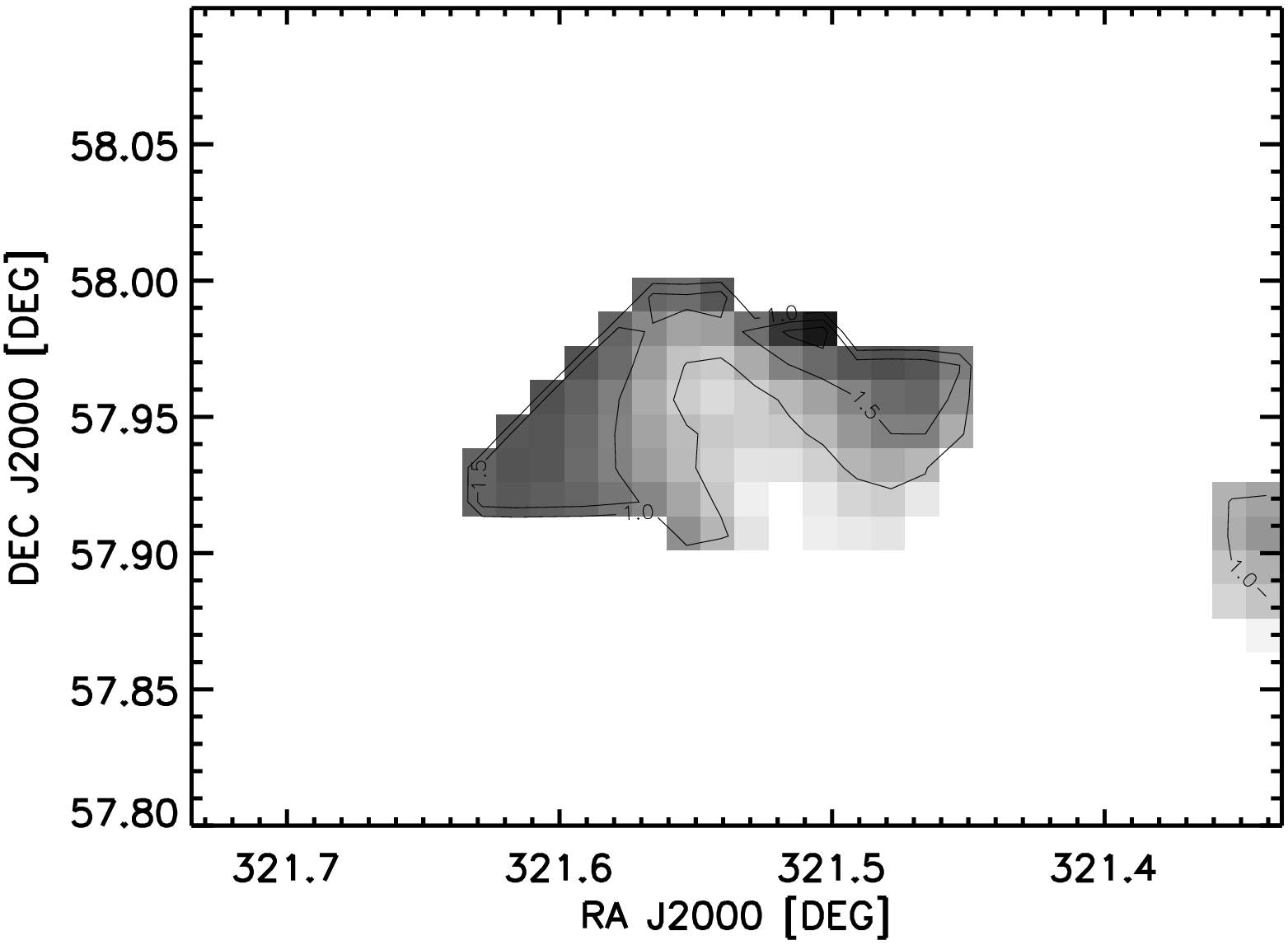}
\hfill
\includegraphics[width=5.8cm,bb=0 0 550 430]{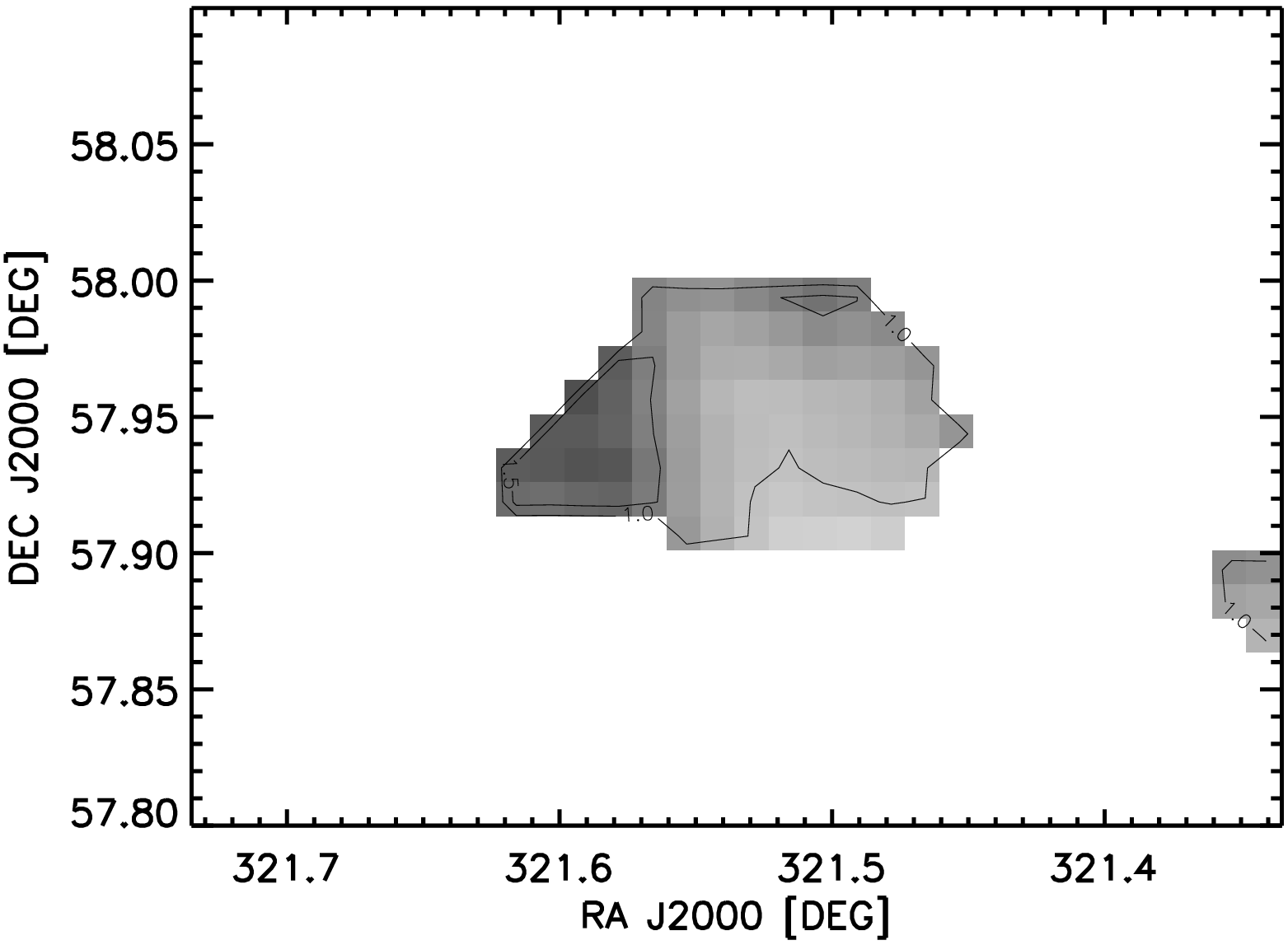}
\hfill
\includegraphics[width=5.8cm,bb=0 0 550 430]{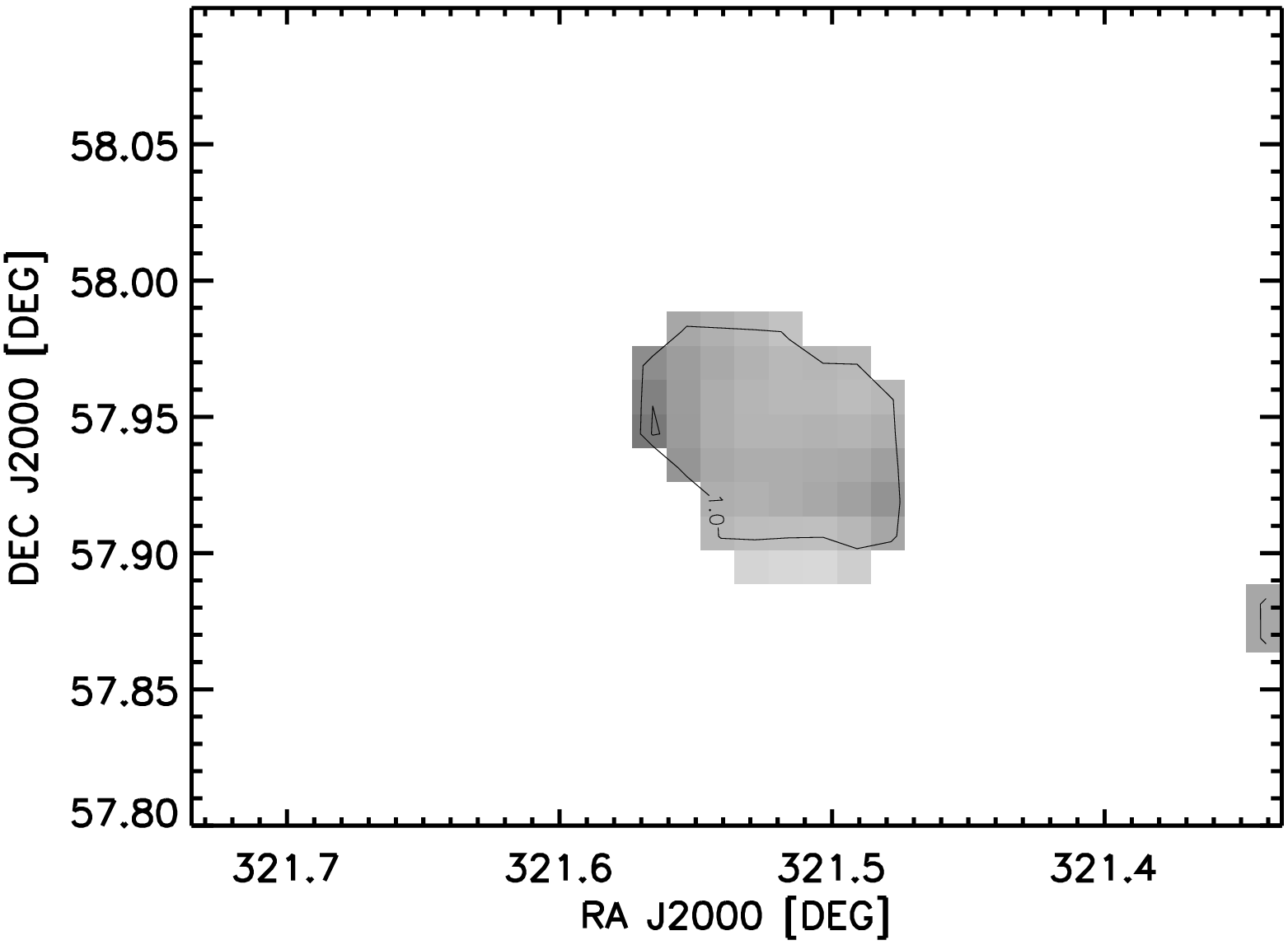}
\caption{\label{ic1396w_beta} {\bf Top:} Gray scale $\beta$-maps obtained from
accumulative star count extinction maps for the wavelength range JH (left), HK
(middle) and JK (right). {\bf Bottom left:} Gray scale $\beta$-map obtained
from $\left< J-H \right>$ and $\left< H-K \right>$ colour excess maps. {\bf
Bottom middle:} $\beta$-map obtained from the combination of $\left< J-H
\right>$ colour excess and J-band star counts. {\bf Bottom right:} Gray scale
$\beta$-map obtained from the combination of $\left< H-K \right>$ colour excess
and H-band star counts. In all panels only areas with extinction or colour
excess values above the 3$\sigma$ noise level are used. Isocontours of
$\beta$=1.0 and 1.5 are marked on gray scale maps where darker colours represent
higher values of $\beta$.}
\end{figure*}

\subsection{$\beta$-maps}

In the top panel of Fig.\,\ref{ic1396w_beta} we show the $\beta$-maps of the
cloud, as determined from the star counts in JH, HK and JK and using
Eq.\,\ref{eq1}. The bottom left panel of Fig.\,\ref{ic1396w_beta} displays the
$\beta$-map obtained from the $\left< J-H \right>$ and $\left< H-K \right>$
colour excess maps and Eq.\,\ref{betacolour}. The two remaining maps are
determined using the combination of star counts, colour excess and
Eq.\,\ref{eq10}. In the bottom middle panel we combine J-band star counts with
$\left< J-H \right>$ colour excess, and in the bottom right panel H-band star
counts and $\left< H-K \right>$ colour excess. In all cases only pixels are
shown, where the used extinction or colour excess is 3$\sigma$ above the noise. 

The first thing to note is that $\beta$-maps obtained using colour excess cover
a larger area, indicating the superiority of the colour excess method to the
star count method in the NIR, in terms of S/N in the extinction maps. We also
see a structure in the $\beta$-map that agrees well with the structure in the
extinction maps. This suggests a dependence of $\beta$ on the column density in
this cloud. In the outer regions of the cloud values for $\beta$ close to the
canonical value for the interstellar medium of 1.85 (Draine
\cite{2003ARA&A..41..241D}) are found. In the centre of the cloud, where
extinction is higher, $\beta$ drops to significantly smaller values, partly
even below 1.0.  When using solely star counts to determine $\beta$ a smaller
area is above the $3 \sigma$ noise level and no structural information can be
obtained. However, the combination of J-band star counts and $\left< J-H
\right>$ colour excess agrees very well with the $\beta$-map obtained solely
from colour excess. The small offset in the absolute $\beta$-values is due the
different control fields used (see discussion above). We note, however, that
differences in $\beta$ within a single cloud can be easily determined by our
method. The correct absolute value for $\beta$ requires a carefully chosen
comparison field, which is free of extinction. 

\subsection{Column density profile}

\begin{figure}
\includegraphics[height=8.5cm,angle=-90,bb=20 30 550 770]{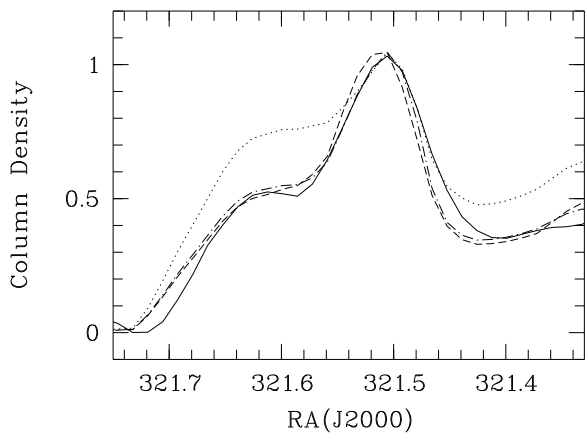}
\caption{\label{ic1396w_profile} Column density profiles (normalised to zero at
the eastern edge of the cloud and unity at the peak) across IC\,1396\,W at a
declination of 57.93$^\circ$. The solid line shows the profile obtained from
J-band star counts and the dotted line is based on $\left< J-H \right>$ colour
excess and a constant $\beta$. The dash-dot line uses a combination of J-band
star counts and $\left< J-H \right>$ colour excess to determine the $\beta$-map,
required to convert $\left< J-H \right>$ into extinction. The dashed line is the
column density profile when using the $\beta$-map obtained from the $\left< J-H
\right>$ and $\left< H-K \right>$ colour excess ratio.}
\end{figure}

How does the change of $\beta$ influence column density profiles of the cloud?
We plot in Fig.\,\ref{ic1396w_profile} the column density profile (normalised to
the peak) from East to West across the cloud at a declination of 57.93$^\circ$.
The solid line shows the profile from J-band star counts and the dotted line
from $\left< J-H \right>$ colour excess (converted with a constant
$\beta$-value). There is a significant difference between the two profiles. The
small peak in the East shows a much higher column density relative to the main
peak when using colour excess instead of star counts. Note that this difference
is unchanged when varying $\beta$, as long as a constant value is used for the
whole cloud. 

We used our $\beta$-map obtained from the combination of J-band star counts and
$\left< J-H \right>$ colour excess to convert the $\left< J-H \right>$-map into
column density. The resulting profile is shown in Fig.\,\ref{ic1396w_profile} as
dot-dashed line. It is obvious that this dramatically changes the column density
profile which now much more closely matches the profile obtained solely from
star counts. Note that we obtain the same result when using the $\beta$-map
obtained from colour excess only (dashed line).

This example shows that the column density structure of clouds can differ
significantly depending on the method used to determine the extinction maps.
This in particularly holds when changes of $\beta$ are found within the cloud.
Thus, extinction maps from accumulative star counts or colour excess (using the
correct $\beta$ values) should be favoured when investigating the
structures/profiles of the denser parts of clouds, where changes of $\beta$ are
more frequently observed. When investigating the outer areas of clouds, where
$\beta$ is very close to the ISM value, or at least constant, colour excess
extinction maps provide the better solution due to their higher signal to
noise. 

\section{Test of extinction determination}

\label{exttest}

Given the possible large uncertainties in the determination of $\beta$, it is
worthy to know the uncertainties of the underlying extinction maps from star
counts or colour excess determination procedures. In particular any systematic
effect should be investigated and corrected for. Here we introduce a scheme to
test the reliability of the extinction determination methods. A detailed
discussion about the influence of the box size to the measured extinction,
systematic offsets in colour excess methods and limitations for reliable
foreground star detection is given.

\subsection{Artificial data}

We used the model of stellar population synthesis of the
Galaxy\footnote{available at http://bison.obs-besancon.fr/modele/} by Robin et
al. \cite{2003A&A...409..523R} to generate artificial data. This model allows us
to generate artificial photometric catalogues for a given region in the sky.
Constraints for the limiting magnitude in each filterband can be applied. We
performed all our simulations using 2MASS catalogue magnitude limits (J$_{\rm
max}$\,=\,16.5, H$_{\rm max}$\,=\,16.0, K$_{\rm max}$\,=\,15.5). As position we
selected ($l$\,=\,81$^\circ$, $b$\,=\,0$^\circ$). This coincides with the
position of the DR\,21 star forming region and hence the quantitative results of
this study can be applied for this area. For any other region in the sky only
the qualitative part of the analysis can be applied; for quantitative results
the simulations described here have to be applied to the specific area.

As mentioned, we determine only the extinction at a certain wavelength when
using accumulative star counts, or we can determine the colour excess between
two wavelengths. The conversion into optical extinction requires the knowledge
of the extinction law. Since we know the extinction law applied to generate the
artificial data (Mathis \cite{1990ARA&A..28...37M}), we are able to directly
compute the optical extinction. This is what finally is used to estimate the
reliability of our extinction determination procedure.

\begin{figure}
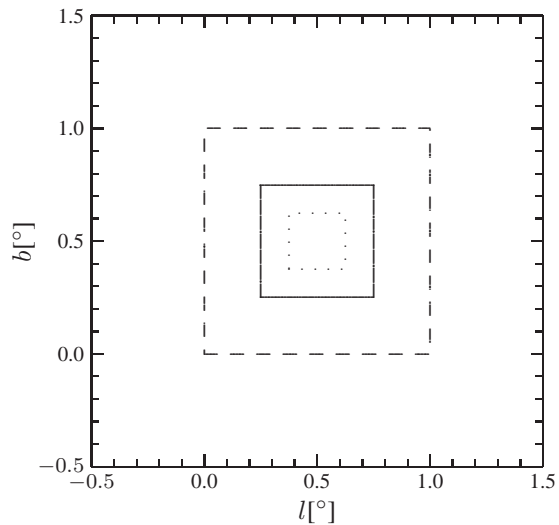

\beginpicture
\setcoordinatesystem units <30mm,30mm> point at 0 0
\setplotarea x from -0.5 to 1.5 , y from -0.5 to 1.5
\axis left label {}
ticks in long numbered from -0.5 to 1.5 by 0.5
      short unlabeled from -0.5 to 1.5 by 0.1 /
\axis right label {}
ticks in long unlabeled from -0.5 to 1.5 by 0.5
      short unlabeled from -0.5 to 1.5 by 0.1 /
\axis bottom label {}
ticks in long numbered from -0.5 to 1.5 by 0.5
      short unlabeled from -0.5 to 1.5 by 0.1 /
\axis top label {}
ticks in long unlabeled from -0.5 to 1.5 by 0.5
      short unlabeled from -0.5 to 1.5 by 0.1 /
\put {\large $l$[$^\circ$]} at 0.5 -0.7
\put {\large \begin{sideways}$b$[$^\circ$]\end{sideways}} at -0.8 0.5
\put {\large \begin{sideways}\end{sideways}} at -1.0 0.5
\setdashes
\plot 0 0 1 0 1 1 0 1 0 0 /
\setdots
\plot 0.375 0.375 0.625 0.375 0.625 0.625 0.375 0.625 0.375 0.375 /
\setsolid
\plot 0.25 0.25 0.75 0.25 0.75 0.75 0.25 0.75 0.25 0.25 /
\endpicture
\caption{\label{testfield} Construction of the test field. An artificial star
catalogue without extinction is generated for the whole 2$^\circ$x2$^\circ$
area. Within a 0.5$^\circ$x0.5$^\circ$ field (solid lines) this catalogue is
exchanged by an artificial catalogue of stars with a cloud at a distance $d$ and
an extinction $A_{\rm V}$. The final extinction map is determined for the
central 1$^\circ$x1$^\circ$ area (dashed lines). We measured the extinction in
the central 0.25$^\circ$x0.25$^\circ$ field (dotted lines).}
\end{figure}

In Figure\,\ref{testfield} we show the layout of our test field, used in the
simulations. The following procedure was applied to generate artificial
observations for this 2$^\circ$x2$^\circ$ field:

\begin{enumerate}

\item A photometric catalogue (set of stars) for a one square degree field
around the coordinates ($l$\,=\,81$^\circ$, $b$\,=\,0$^\circ$) was downloaded
using the 2MASS magnitude limits, a standard galactic extinction law of $A_{\rm
V}$\,=\,0.7\,mag$\cdot$kpc$^{-1}$ and no clouds. To fill the 2$^\circ$x2$^\circ$
field with this set of stars, each star was supplied with four different,
artificial, uniformly distributed coordinates, that placed it in the
2$^\circ$x2$^\circ$ field. In other words, these artificial coordinates are
pairs of random numbers generated to mimic a uniform distribution of stars in
the sky.

\item A photometric catalogue with the same magnitude limits and position was
downloaded, but this time including a cloud with a certain distance and a
certain extinction. Each star in this second catalogue was supplied with
artificial, uniformly distributed coordinates, that placed it in the central
1$^\circ$x1$^\circ$ field. 

\item The two catalogues where merged in a way that all stars in the first
catalogue which fell in the central 0.5$^\circ$x0.5$^\circ$ area (solid lines in
Fig.\,\ref{testfield}) are replaced by the stars of the second catalogue in this
area. In other words we created an artificial catalogue, simulating a squared
0.5$^\circ$x0.5$^\circ$ cloud in the centre of a 2$^\circ$x2$^\circ$ field.

\end{enumerate}

For each of these artificial datasets we determined the extinction map for the
central 1$^\circ$x1$^\circ$ field and measured the mean and median extinction of
the cloud in the central 0.25$^\circ$x0.25$^\circ$.  This set-up allows us to
perform different tests to determine the reliability of our extinction
determination method. 

\subsection{Zero extinction (i.e. no clouds)}

\label{noclouds}

Here we consider an object with zero extinction at zero distance to estimate the
contribution of the simulated photometric datasets and artificial coordinates to
the noise of the method. Such an object is equivalent to consider "no cloud",
but it is also representative of a region with negligible extinction.

We first fix the set of stars in the control field, the cloud and the
coordinates assigned to these stars. This allows us to determine how the chosen
box size influences the noise in the extinction map. As expected we find that
the box size is inversely proportional to the noise, with very small deviations.
This can be explained since the noise should be inversely proportional to the
square root of the number of stars in each box, and then inversely proportional
to the box size. As box size for all subsequent analysis we chose 1.2', which
ensures on average 25 stars per box.

\begin{table}
\centering
\caption{\label{noisetable} Contribution of artificial coordinates and set of
stars on the noise in the extinction maps from star counts and colour excess for
model clouds with negligible extinction. We show here results for K-band star
counts and $\left< H-K \right>$ colour excess methods. The noise is given in
magnitudes of $A_{\rm V}$.}
\begin{tabular}{lcc}
 & \multicolumn{2}{c}{noise from} \\ 
influences & star counts & colour excess \\ 
\noalign{\smallskip}
\hline
\noalign{\smallskip}
artificial coordinates & 0.15 & 0.03 \\
set of stars in & & \\
\hspace{1cm} control field & 0.13 & 0.05 \\
\hspace{1cm} cloud & 0.10 & 0.04 \\
\noalign{\smallskip}
\hline
\noalign{\smallskip}
total noise & 0.22 & 0.07 \\
\end{tabular}
\end{table}

We have performed several tests in order to disentangle the different components
contributing to the noise in the extinction maps. In particular: a) the
artificial coordinates of the stars in the field; b) the set of stars in the
control field; c) the set of stars in the area of the cloud. The contribution to
the noise from each of those three components can be determined from the scatter
in the synthetic extinction maps, by varying one component and fixing the other
two. 

Our results are summarised in Table\,\ref{noisetable}. There we compare the
noise (in magnitudes of $A_{\rm V}$, converted using the extinction law of
Mathis \cite{1990ARA&A..28...37M}) between K-band star counts and $\left< H-K
\right>$ colour excess. We find that the star count method leads to three times
larger noise (for our chosen magnitude limits and box size). When using the
colour excess method, the set of stars in the control field or the cloud is
responsible for most of the scatter. For star counts, however, the artificial
coordinates dominate the noise. 

\subsection{Clouds at zero distance}

Since we have established the influence of the simulated star sets and
artificial coordinates, we now investigate the changes for different cloud
extinction values ($A_{\rm V}$ = 0.0, 0.2, 0.5, 1.0, 3.0, 5.0, 7.0, 10.0\,mag).
To avoid uncertainties due to foreground stars, the model clouds are placed at
zero distance.

\begin{figure}
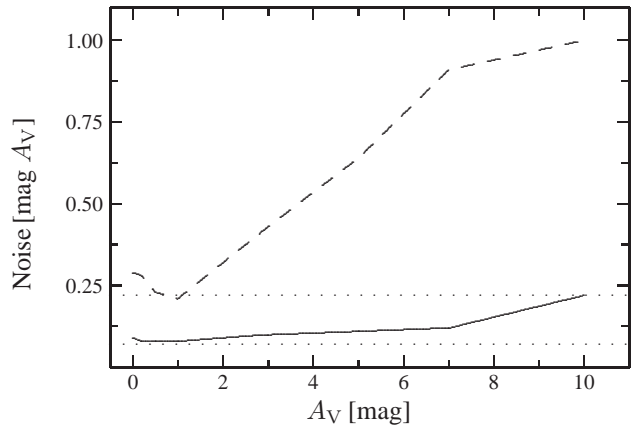

\beginpicture
\setcoordinatesystem units <6mm,43.5mm> point at 0 0
\setplotarea x from -0.5 to 11 , y from -0 to 1.1
\axis left label {}
ticks in long numbered from 0.25 to 1.0 by 0.25
      short unlabeled from 0. to 1.1 by 0.125 /
\axis right label {}
ticks in long unlabeled from 0.25 to 1.0 by 0.25
      short unlabeled from 0. to 1.1 by 0.125 /
\axis bottom label {}
ticks in long numbered from 0 to 10 by 2
      short unlabeled from 0 to 10 by 1 /
\axis top label {}
ticks in long unlabeled from 0 to 10 by 2
      short unlabeled from 0 to 10 by 1 /
\put {\large $A_{\rm V}$\,[mag]} at 5 -0.14
\put {\large \begin{sideways}Noise\,[mag $A_{\rm V}$]\end{sideways}} at -2.4 0.5
\setsolid
\plot 0 0.09 0.2 0.08 0.5 0.08 1 0.08 3 0.1 5 0.11 7 0.12 10 0.22 /
\setdashes
\plot 0 0.29 0.2 0.28 0.5 0.23 1 0.21 3 0.43 5 0.64 7 0.91 10 1 /
\setdots
\plot -0.5 0.22 11 0.22 /
\plot -0.5 0.07 11 0.07 /
\endpicture
\caption{\label{noiseav} Noise in the extinction maps as a function of $A_{\rm
V}$ of our model clouds. The solid line results from the colour excess method
$\left< H-K \right>$ and the dashed line from accumulated star counts (K-band).
The two dotted horizontal lines mark the noise due to the set of simulated stars
and artificial coordinates determined by our test in Sect.\,\ref{noclouds} for
the two methods (the upper line is for the K-band star counts, the lower for the
) $\left< H-K \right>$ colour excess.}
\end{figure}

We test how the scatter in the extinction map changes with $A_{\rm V}$.
Figure\,\ref{noiseav} shows how the noise in the extinction maps changes with
$A_{\rm V}$ of the model cloud. The dashed line corresponds to the colour excess
method and the dotted line to the accumulated star counts. For low extinction
values, below one mag of $A_{\rm V}$, the measured values agree very well with
the estimated values in the above section. This can be explained since such low
extinction values only marginally influence the number of stars in each box and
therefore the noise in the maps. For larger $A_{\rm V}$-values, however, the
scatter increases significantly. This is largely due to the fact that a much
lower number of stars is present in each box. When comparing the two methods we
find that accumulated star counts lead on average to a three to five times
higher noise.

\subsection{Systematic offsets --- distance correction}

\label{distcorr}

\begin{figure}
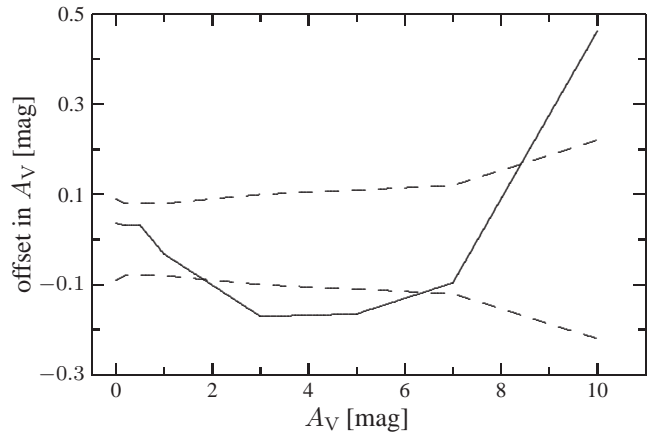

\beginpicture
\setcoordinatesystem units <6.4mm,60mm> point at 0 0
\setplotarea x from -0.5 to 11 , y from -0.3 to 0.5
\axis left label {}
ticks in long numbered from -0.3 to 0.5 by 0.2
      short unlabeled from -0.3 to 0.5 by 0.1 /
\axis right label {}
ticks in long unlabeled from -0.3 to 0.5 by 0.2
      short unlabeled from -0.3 to 0.5 by 0.1 /
\axis bottom label {}
ticks in long numbered from 0 to 10 by 2
      short unlabeled from 0 to 10 by 1 /
\axis top label {}
ticks in long unlabeled from 0 to 10 by 2
      short unlabeled from 0 to 10 by 1 /
\put {\large $A_{\rm V}$\,[mag]} at 5 -0.4
\put {\large \begin{sideways}offset in $A_{\rm V}$\,[mag]\end{sideways}} at -1.9 0.1
\setsolid
\plot 0 0.036 0.2 0.032 0.5 0.032 1 -0.032 3 -0.17 5 -0.165 7 -0.096 10 0.462 /
\setdashes
\plot 0 0.09 0.2 0.08 0.5 0.08 1 0.08 3 0.1 5 0.11 7 0.12 10 0.22 /
\plot 0 -0.09 0.2 -0.08 0.5 -0.08 1 -0.08 3 -0.1 5 -0.11 7 -0.12 10 -0.22 /
\endpicture
\caption{\label{offset} Systematic offsets (solid line) of the recovered
extinction values for our model clouds depending on the optical extinction when
using the colour excess method. The dashed lines mark the $\pm 1\sigma$
uncertainties.}
\end{figure}

We simulated clouds with different extinction values at zero distance. This
ensures that no foreground stars are present and we should be able to fully
recover the applied extinction values of our artificial clouds, given the fact
that we know the reddening law applied to the artificial data (Mathis
\cite{1990ARA&A..28...37M}). Using accumulated star counts the measured
extinction is in very good agreement with the real (input) value. The measured
values scatter only 0.4\,$\sigma$ around the input value and hence no measurable
systematic effects exist. In absolute terms these deviations are larger than
what is obtained by the colour excess method, due to the in generally higher
noise of the accumulated star count technique. The offsets obtained by the
colour excess method are hence small but systematic, i.e. above the internal
scatter of the method. Fig.\,\ref{offset} shows the variation of the offset with
$A_{\rm V}$. In the following we discuss their cause.

For small extinction values ($A_{\rm V} < 1$\,mag) the deviations are smaller
than 1\,$\sigma$. For higher extinction ($A_{\rm V}$ up to 7\,mag) we
systematically underestimate the extinction by up to 2$\sigma$. For very high
column densities ($A_{\rm V} \approx 10$\,mag) this effect turns and we
overestimate the extinction.  Clearly an explanation for this effect is needed.
Due to the setup of our test, a contribution from foreground stars can be
neglected. For small values of extinction (below 1\,mag $A_{\rm V}$) the effect
is negligible. This implies the reddening of the stars due to the dust in the
cloud itself is the cause of the offsets. When comparing the mean/median colour
of stars in the cloud and an extinction free control-field, we have to keep in
mind that in both areas stars down to a certain limiting magnitude are detected.
We see, however, in these two fields different populations of stars. Stars
behind the cloud, detected down to the same limiting magnitude, are on average
at a different distance and redder. Thus the mean/median intrinsic colour of
stars behind the cloud differs from the mean/median intrinsic colour of stars in
the control field. In conclusion {\it all methods using colour excess} to
estimate extinction {\it are subject to this effect}. Even if small for low
extinction values it might become significant and systematic for denser regions,
changing effectively the clouds column density profile. 

 %
 %

The particular quantitative behaviour of the offsets is a complex interplay of
various parameters. The position in the sky and the limiting magnitude of the
observations determine the population of stars contained in the control field.
The extinction of the cloud then determines how this population is changed when
seen through the cloud. Extinction could have the effect that less red-giant
background stars are seen through the cloud. Also it is possible that less faint
stars close behind the cloud are visible. If one of those effects prevails,
systematic offsets are expected. The extinction further could cause the median
distance of the stars seen through the cloud be different from the the stars in
the control field. This can also introduce systematic offsets due to
interstellar extinction (typically 0.7\,mag $A_{\rm V}$/kpc, but reaching up to
2\,mag/kpc in certain regions; Joshi \cite{2005MNRAS.362.1259J}). All these
effects are further influenced by the actual distance of the cloud. Due to this
complexity it is not possible to describe the offsets using a simple model with
a few parameters. The only way to determine the quantitative offsets and to
correct them is to perform simulations similar to the ones presented here, using
the limiting magnitude and coordinates of the real observations as well as the
distance and extinction range for the observed cloud. We note that the
quantitative effect most likely depends on the particular model used. However,
it is important to realise its existence and to know its possible order of
magnitude. Thus it should be considered in the analyses of high extinction
regions with colour excess methods.

Do those offsets have a systematic influence on the determination of
$\beta$? One can estimate that the value of $\beta$, determined with or without
the correction of the systematic offset, differs by at most 5\,\%. This is
usually much smaller than the scatter of $\beta$ (in the order of 30\,\% or
higher) due to the scatter in the colour excess maps (see e.g.
Fig.\,\ref{err_bet}). Only for very high S/N data with a very small internal
scatter of $\beta$ these offsets become important.

\subsection{Clouds with variable distance}

Foreground stars have to be identified and removed from the input catalogue
because they influence the correct extinction determination. This is usually done
by identifying all stars in the field of the cloud that are much bluer than what
would be expected from the extinction of the cloud. The notion is stars behind
the cloud are red, stars in front of the cloud are blue. This requires, however,
the detection of a significant extinction towards the cloud prior to the
foreground star identification. Here we investigate the difficulties for the
identification of foreground stars for the star counting and colour excess
method.

\begin{figure}
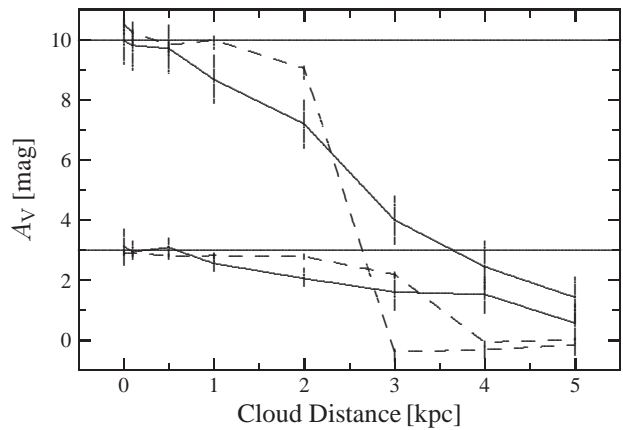

\beginpicture
\setcoordinatesystem units <12mm,4mm> point at 0 0
\setplotarea x from -0.5 to 5.5 , y from -1 to 11
\axis left label {}
ticks in long numbered from -0 to 10 by 2
      short unlabeled from -0 to 10 by 1 /
\axis right label {}
ticks in long unlabeled from -0 to 10 by 2
      short unlabeled from -0 to 10 by 1 /
\axis bottom label {}
ticks in long numbered from 0 to 5 by 1
      short unlabeled from 0 to 5 by 0.5 /
\axis top label {}
ticks in long unlabeled from 0 to 5 by 1
      short unlabeled from 0 to 5 by 0.5 /
\put {\large Cloud Distance\,[kpc]} at 2.5 -2.5
\put {\large \begin{sideways}$A_{\rm V}$\,[mag]\end{sideways}} at -1.1 5
\setsolid
\plot -0.5 10 5.5 10 /
\plot 0 9.98 0.1 9.81 0.5 9.71 1 8.68 2 7.20 3 4.01 4 2.45 5 1.43 /
\plot 0 9.2 0 10.8 /
\plot 0.1 9.0 0.1 10.6 /
\plot 0.5 8.9 0.5 10.5 /
\plot 1 7.9 1 9.5 /
\plot 2 6.4 2 8.0 /
\plot 3 3.2 3 4.8 /
\plot 4 1.7 4 3.3 /
\plot 5 0.6 5 2.1 /
\setdashes
\plot 0 10.52 0.1 10.24 0.5 9.83 1 9.99 2 9.05 3 -0.37 4 -0.31 5 -0.15 /
\plot 0 10.2 0 10.8 /
\plot 0.1 9.9 0.1 10.5 /
\plot 0.5 9.5 0.5 10.1 /
\plot 1 9.7 1 10.3 /
\plot 2 8.7 2 9.3 /
\plot 3 -0.7 3 -0.1 /
\plot 4 -0.6 4 0 /
\plot 5 -0.5 5 0.1 /
\setsolid
\plot -0.5 3 5.5 3 /
\plot 0 3.13 0.1 2.95 0.5 3.09 1 2.56 2 2.06 3 1.61 4 1.53 5 0.58 /
\plot 0 2.5 0 3.7 /
\plot 0.1 2.7 0.1 3.3 /
\plot 0.5 2.8 0.5 3.4 /
\plot 1 2.3 1 2.9 /
\plot 2 1.8 2 2.4 /
\plot 3 1 3 2.2 /
\plot 4 0.9 4 2.1 /
\plot 5 0 5 1.2 /
\setdashes
\plot 0 2.90 0.1 2.92 0.5 2.80 1 2.82 2 2.79 3 2.20 4 -0.06 5 0.03 /
\plot 0 2.8 0 3 /
\plot 0.1 2.8 0.1 3 /
\plot 0.5 2.7 0.5 2.9 /
\plot 1 2.7 1 2.9 /
\plot 2 2.7 2 2.9 /
\plot 3 2.1 3 2.3 /
\plot 4 -0.2 4 0 /
\plot 5 -0.1 5 0.1 /
\setsolid
\endpicture
\caption{\label{distance} Recovered extinction depending on the distance of the
cloud using accumulative star counts (solid) and colour excess (dashed) methods
for clouds with an optical extinction of 10 (lines starting at 10\,mag for zero
distance) and 3\,mag (lines starting at 3\,mag for zero distance). The
horizontal lines mark the input cloud extinction.} 
\end{figure}

We analyse how the extinction of our model cloud depends on the distance of the
cloud. Distances of 0, 0.1, 0.5, 1, 2, 3, 4 and 5\,kpc were considered. In
Fig.\,\ref{distance} we plot how the measured extinction depends on the cloud
distance for a 10 and 3\,mag $A_{\rm V}$ cloud. Accumulative star counts (solid
lines) and colour excess (dashed lines) extinction maps, prior to foreground
stars selection, are shown.

There are obvious differences between the two methods. The star counting method
shows a steady decline of the measured extinction with increasing cloud
distance. For the colour excess method at first only a very small decline with
cloud distance is observed. After a certain critical distance, however, the
measured extinction drops to zero. This is the distance at which more than half
of the stars in the field are foreground to the cloud, and the median colour
represent foreground stars. The higher the extinction in the cloud the smaller
is the maximum distance up to which a cloud can be detected using colour excess
since less background stars are detected in the field. 

When investigating distant clouds, it is hence of advantage to use the star
count extinction maps for foreground star identification. In our example in
Fig.\,\ref{distance} it is observed that we are not able to identify foreground
stars in a 10\,mag $A_{\rm V}$ cloud with a distance larger than 2\,kpc when
using colour excess extinction maps. Conversely, the star counting technique is
able to detect this cloud at distances of up to 4\,kpc at a 3\,$\sigma$ level.
For 3\,mag $A_{\rm V}$ the critical distance for the colour excess method is
3\,kpc, compared to 4\,kpc for the star counting technique. 

\subsection{Suggested extinction determination method}

Given the results presented here, we suggest the following method to determine
extinction maps:

\begin{enumerate}

\item Star count and colour excess extinction maps are calculated without
foreground star corrections. These maps are used to determine the opacity index
$\beta$, which is used to convert the colour excess maps into extinction maps. 

\item Foreground stars have to be selected. Extinction maps from colour excess
are usually less noisy but are less reliable for distant clouds. Thus the
extinction map that possesses the larger extinction values should be used for
the foreground star selection.

\item Now one needs to iterate the first two steps until no further foreground
stars are found. Foreground stars cannot be identified in regions without
extinction. We thus have to consider the accumulated star count diagram of the
foreground stars when determining the extinction map using star counts. Using
the colours of the identified foreground stars and Monte-Carlo techniques, a
colour offset for background regions can be determined.

\item The final star count and colour excess maps are used to determine $\beta$
and to convert the colour excess into extinction maps.

\item For a most accurate calibration determine the distance correction using
the cloud distance and coordinates, as well as the completeness limit of the
observations.

\end{enumerate}

\section{Conclusions}

\label{conclusions}

\begin{itemize}

\item We have explained {\it how} a map of the near infrared extinction powerlaw
index can be determined from colour excess, star counting, or a combination of
both. A detailed analysis of the uncertainties of each method is performed. We
find that generally high S/N extinction measurements are required for a
precision determination of $\beta$. However, taking into account the cloud
structure and possible dependence of $\beta$ on the column density, allows us to
average different measurements and increase the accuracy. 

\item We find that for 2MASS data of IC\,1396\,W the combination of star
counting and colour excess methods lead to the highest signal to noise
$\beta$-map. We find a dependence of the NIR extinction powerlaw index on the
column density in the cloud. In low column density regions $\beta$ is in
agreement with the canonical value for the interstellar medium of 1.85 (Draine
\cite{2003ARA&A..41..241D}). At higher column densities, however, $\beta$ drops
to values as low as 1.0. We further show that the column density profile of the
cloud differs when measured by colour excess or star count extinction maps.
Converting the colour excess into extinction using the correct value of $\beta$
at each position, corrects these differences. It is thus vital to check and
correct for any systematic changes of $\beta$ in dark clouds, prior to analysing
the column density profile in extinction maps obtained from colour excess.

\item We present a scheme to test the accuracy of extinction determination
methods using artificial observational data. This is based on the model of
stellar population synthesis of the Galaxy by Robin et al.
\cite{2003A&A...409..523R}. We show that all extinction determination methods
based on colour excess are subject to small but systematic offsets. These are
due to the fact that the population of stars seen through a cloud is different
from the population of stars in an extinction free control field (even if close
by). The resulting offsets, however, can be determined with the knowledge of the
cloud position, distance and extinction, as well as the completeness limit of
the photometric catalogue. The same scheme allows us to estimate the distance
out to which clouds can be detected with star counts and colour excess.
Generally, star counts allow us to detect more distant clouds than colour
excess. This is particularly useful for the first step of foreground star
selection in the case of distant clouds.

\end{itemize}

\section*{Acknowledgements}
 
We would like to thank the anonymous referee for significant comments to
improve the paper. D.\,F. received financial support from the Cosmo-Grid
project, funded by the Program for Research in Third Level Institutions under
the National Development Plan and with assistance from the European Regional
Development Fund. This publication makes use of data products from the Two
Micron All Sky Survey, which is a joint project of the University of
Massachusetts  and the Infrared Processing and Analysis Center/California
Institute of  Technology, funded by the National Aeronautics and Space
Administration and the National Science Foundation.

\label{lastpage}

\end{document}